\documentclass[11pt]{article}
\pdfoutput=1

\usepackage{euscript}
\usepackage{amssymb}
\usepackage{amsfonts}
\usepackage{amsbsy}
\usepackage{epsfig}
\usepackage{amsthm}
\usepackage{amscd}
\usepackage{amstext}
\usepackage{verbatim}
\usepackage{amsmath}
\usepackage{cancel}
\usepackage{capt-of}
\usepackage{empheq}
\usepackage{subfigure}
\usepackage{xcolor}

\usepackage{cite}
\usepackage{bm}

\usepackage[pdftex,linktoc=all]{hyperref}
\hypersetup{ 
colorlinks=true, 
linkcolor=black, 
citecolor=red, 
}

\def\ben{\begin{equation}}
\def\een{\end{equation}}

\let\a=\alpha   \let\d=\delta 
   \let\k=\kappa

\let\w=\omega

\let\pa=\partial
\def\be{\begin{equation}}
\def\ee{\end{equation}}
\def\beq{\begin{equation}}
\def\eeq{\end{equation}}
\def\ba{\begin{array}}
\def\ea{\end{array}}

\def\dalemb#1#2{{\vbox{\hrule height .#2pt
       \hbox{\vrule width.#2pt height#1pt \kern#1pt
               \vrule width.#2pt}
       \hrule height.#2pt}}}

\newcommand{\bea}{\begin{eqnarray}}
\newcommand{\eea}{\end{eqnarray}}

\makeatletter
\newcommand*\bigcdot{\mathpalette\bigcdot@{.5}}
\newcommand*\bigcdot@[2]{\mathbin{\vcenter{\hbox{\scalebox{#2}{$\m@th#1\bullet$}}}}}
\makeatother

\renewcommand{\eqref}[1]{(\ref{#1})}

\def\H{{{\rm H}}}

\def\Im{{{\frak{Im}}}}
\def\Re{{{\frak{Re}}}}

\def\Lag{{\mathcal{L}}}

\def\ocal{{\mathcal{O}}}

\def \d {\partial}
\renewcommand{\Im}[0]{\operatorname{Im}}
\renewcommand{\Re}[0]{\operatorname{Re}}

\numberwithin{equation}{section}

\begin{document}
\frenchspacing
\begin{center}

{ \Large {\bf
Theory of the collective magnetophonon resonance and \\
melting of the field-induced Wigner solid
}}

\vspace{1cm}

Luca V. Delacr\'etaz$^{1}$, Blaise Gout\'eraux$^{2,3}$, Sean A. Hartnoll$^{1,4}$ and Anna Karlsson$^{5,6}$

\vspace{1cm}

{\small
$^{1}${\it Department of Physics, Stanford University, \\
Stanford, CA 94305-4060, USA }}

\vspace{0.3cm}

{\small
$^{2}${\it
Center for Theoretical Physics, \'Ecole Polytechnique, CNRS UMR 7644, \\
Universit\'e Paris-Saclay, 91128, Palaiseau, France}}

\vspace{0.3cm}

{\small
$^{3}${\it
Nordita, KTH Royal Institute of Technology and Stockholm University, \\
Roslagstullsbacken 23, SE-106 91 Stockholm, Sweden}}

\vspace{0.3cm}

{\small
$^{4}${\it
Stanford Institute for Materials and Energy Science, SLAC National Accelerator Laboratory, \\
2575 Sand Hill Road, Menlo Park,
CA 94025, USA}}

\vspace{0.3cm}

{\small
$^{5}${\it
Institute for Advanced Study, School of Natural Sciences, \\
1 Einstein Drive, Princeton, NJ 08540, USA}}

\vspace{0.3cm}

{\small
$^{6}${\it
Division for Theoretical Physics, Department of Physics, \\
Chalmers University of Technology, SE-412 96 Gothenburg, Sweden}}

\vspace{1cm}

\end{center}

\begin{abstract}

Electron solid phases of matter are revealed by characteristic vibrational resonances. Sufficiently large magnetic fields can overcome the effects of disorder, leading to a weakly pinned collective mode called the magnetophonon. Consequently, in this regime it is possible to develop a tightly constrained hydrodynamic theory of pinned magnetophonons. The behavior of the magnetophonon resonance across thermal and quantum melting transitions has been experimentally characterized in two-dimensional electron systems. Applying our theory to these transitions we explain several key features of the data: (i) violation of the Fukuyama-Lee sum rule as the transition is approached is directly tied to the non-Lorentzian form taken by the resonance and (ii) the non-Lorentzian shape is caused by characteristic dissipative channels that become especially important close to melting: proliferating dislocations and uncondensed charge carriers.

\end{abstract}
\thispagestyle{empty}
\pagebreak
\pagenumbering{arabic}

\newpage


\tableofcontents
\noindent\hrulefill

\section{Introduction}

Magnetophonons are vibrational modes of electron solid phases of matter in the presence of a magnetic field \cite{PhysRevB.15.1959}.
A remarkable fact about magnetophonons is that their long wavelength modes can survive at low energies in the presence of disorder. Without a magnetic field, the sound modes of electronic translational order are typically pinned to a high microscopic frequency scale $\omega_o$ by disorder \cite{RevModPhys.60.1129}.
However, a magnetic field hybridizes the longitudinal and shear sound modes into the so-called magnetophonon and magnetoplasmon. At large field, the magnetoplasmon tends towards the high cyclotron frequency $\omega_c$, while the magnetophonon becomes parametrically light, with peak frequency $\omega_\text{pk} \sim \omega_o^2/\omega_c$ \cite{PhysRevB.18.6245}.

The universal low energy, long wavelength excitations of any thermal medium are described by hydrodynamics \cite{chaikin1995principles}. For sufficiently pure samples (small $\omega_\text{pk}$) the magnetophonon resonance will be in this collective regime, and this limit provides a well-defined theoretical starting point. Such low energy pinned magnetophonons will have their own hydrodynamic theory, consistently decoupled from non-universal high energy dynamics. While magnetophonon modes have been investigated for decades from a microscopic perspective, a systematic hydrodynamic theory has not been formulated. Upon formulating this theory, we will be able to shed light on several open problems concerning the observed melting behavior of magnetophonon resonances.

In sufficiently large magnetic fields, the ground state of two-dimensional electrons systems such those arising in GaAs/GaAlAs heterostructures is expected to be a Wigner solid \cite{doi:10.1002/9783527617258.ch9, refId0}. Re-entrant insulating phases between quantum Hall plateaux \cite{PhysRevLett.65.633, PhysRevB.44.8107} are also naturally interpreted as Wigner solids, as evidenced by threshold behavior in their nonlinear conductivity \cite{PhysRevLett.65.2189,PhysRevLett.66.3285,PhysRevB.44.8107}. The direct detection of a pinned magnetophonon resonance, however, has long been considered the smoking gun signature of crystalline order in a large field \cite{PhysRevLett.60.2765,PhysRevB.45.11342,PhysRevB.45.13784, PhysRevLett.66.3285, PhysRevLett.79.1353,ENGEL1997167,ENGEL1997111,HENNIGAN199853,PhysRevB.61.10905}. More recent measurements have systematically investigated the form and location of the long wavelength magnetophonon resonance as a function of filling fraction, electron density, disorder and temperature \cite{PhysRevLett.89.176802,PhysRevLett.93.206805, Chen2006,doi:10.1142/S0217979207042860, PhysRevB.89.075310,PhysRevB.92.035121}, and have also found collective vibrational modes in the immediate vicinity of quantum Hall phases \cite{PhysRevLett.91.016801, PhysRevLett.105.126803, PhysRevB.95.045417}.

Aspects of the observed dependence of the magnetophonon peak on parameters such as field, density and disorder have been successfully described by microscopic theories of harmonic lattice vibrations in the presence of disorder and Coulomb interactions \cite{PhysRevB.18.6245,PhysRevB.46.3920, PhysRevB.59.2120, PhysRevLett.80.3827, PhysRevB.65.035312, PhysRevB.62.7553}. For an overview of these results, see \cite{chen2005quantum}. Little is understood about the effect of many-body interactions on magnetophonon dynamics, although these are likely important for the melting dynamics of the Wigner solid \cite{Chen2006}. Indeed, the behavior of the magnetophonon resonance as the solid melts is at odds with current theory \cite{chen2005quantum}.

Our collective approach describes a limit of extreme dominance of interactions, formally opposite to the harmonic vibration regime. Specifically, hydrodynamics is valid at the lowest energy scales $\omega \ll 1/\tau_\text{eq}$ with $\tau_\text{eq}$ the local thermal equilibration time. In this limit almost everything has decayed, and one need only keep track of the dynamics of a finite number of conserved densities and Goldstone modes, leading to a robust and powerfully constrained structure for the magnetophonon resonance, described in \S \ref{sec:hydro}. Functions of frequency are analytic and dissipation is described by a finite number of transport coefficients. Specifically, we find the peak in the optical conductivity takes the form
\be\label{eq:sintro}
\sigma(\omega)= \nu \, \omega_\text{pk}\frac{(1-a^2)(-i\omega+\Omega)-2a\omega_\text{pk}}{(-i\omega+\Omega)^2+\omega_\text{pk}^2}\,.
\ee
Here $\nu$ is the filling fraction. In addition to $\omega_\text{pk}$ there are only two undetermined coefficients characterizing the peak, $\Omega$ and $a$. We will find that this functional form gives a good fit to the data across the entire parameter range that we consider, see Fig. \ref{fig:plots} below.
The simple, analytic-in-frequency form (\ref{eq:sintro}) is the key sense in which our theory is collective: the conductivity is determined by only a small amount of microscopic data, packaged into three coefficients. In practice this can be quite a weak requirement. For example, the low-frequency response of conventional metals is often well-described by the Drude form, which is analytic and has only two undetermined coefficients, despite these metals not being at all hydrodynamic in the sense of momentum being longer lived than all other modes.

The peak width $\Omega$ has been widely considered. The constant $a$ controls the non-Lorentzian shape of the peak, and is also crucial. In particular, the spectral weight in the peak is
\be\label{eq:S}
S = {\textstyle \frac{\pi}{2}} (1-a^2) \nu \, \omega_\text{pk} \,.
\ee
This expression recovers the well-known Fukuyama-Lee result \cite{PhysRevB.18.6245} when $a = 0$. We will find that fits to the data require $a$ to be nonzero and to increase dramatically as the solid melts as a function of temperature or filling. In this way we will quantitatively explain violations of the Fukuyama-Lee sum rule that have been previously noted --- they are directly tied (by hydrodynamics) to the non-Lorentzian shape of the peak \eqref{eq:sintro}. See Fig. \ref{fig:FL} below.

The coefficients $a$ and $\Omega$ can be evaluated using Kubo formulae, that we derive and evaluate in \S \ref{sec:dis}. Disorder gives a contribution $\Omega \sim \omega_\text{pk}$ to the width, somewhat analogous to the results obtained in microscopic theories \cite{PhysRevB.18.6245, PhysRevB.59.2120, PhysRevLett.80.3827, PhysRevB.65.035312, PhysRevB.62.7553}. An especially universal contribution to $a$ --- loosely, from dissipation of the pinned phase into currents of uncondensed carriers --- gives $a \propto \Omega/\omega_\text{pk}$.
A further source of phase relaxation are mobile dislocations \cite{PhysRevLett.66.652}. In the pinned regime these are also found to lead to $a \propto \Omega/\omega_\text{pk} \sim x$, the density of mobile dislocations.
In this way we obtain two mechanisms that suggest that in a phase-disordering melting transition, driven by a rapid increase in $\Omega/\omega_\text{pk}$, the constant $a$ should also increase. This is precisely what is seen in several of our fits to the data. See Figs. \ref{fig:ABomegas} and \ref{fig:FL} below. Taking all the above together, we obtain a physically plausible and quantitatively accurate picture of the thermal and quantum melting dynamics observed in \cite{Chen2006,chen2005quantum}.

\section{Magnetophonon hydrodynamics}
\label{sec:hydro}

In the absence of a magnetic field, incommensurate translational order leads to a Goldstone mode for each spontaneously broken translation.
These Goldstone fields produce a new sound mode -- shear sound -- in addition to the usual longitudinal sound present in translation-invariant systems \cite{chaikin1995principles}. Both sound modes are gapped (or `pinned') when translations are explicitly broken by e.g. disorder. If pinning is sufficiently weak these modes can still be described within hydrodynamics. The modes are now called pseudo-Goldstone bosons. Pinned sound modes are responsible for a peak in the dynamical response at nonzero frequency $\omega_o$ \cite{RevModPhys.60.1129}. Often, however, $\omega_o$ is large and outside of the hydrodynamic regime.

A large magnetic field hybridizes the longitudinal and shear sound modes into so-called magnetophonon and magnetoplasmon modes, with gaps of order $\omega_o^2/\omega_c$ and $\omega_c$ respectively, where $\omega_c = n B/\chi_{\pi\pi}$ is the cyclotron frequency \cite{PhysRevB.18.6245}. Here $\chi_{\pi\pi}$ is the momentum susceptibility; for example, in a Galilean-invariant system $\chi_{\pi\pi} = n m$ is the mass density. In the experimentally relevant limit $\omega_o \ll \omega_c$ the magnetophonon becomes light. Hydrodynamics applies at frequencies below the local thermal equilibration rate $1/\tau_{\rm eq}$. Therefore, if
\begin{equation}\label{eq_regime}
\omega_o^2/\omega_c \ \ll \ 1/\tau_{\rm eq} \ \ll \  \omega_c \, ,
\end{equation}
then there should exist a hydrodynamic theory of the magnetophonon alone, without the high-energy magnetoplasmon. This theory will hold even if the pinning frequency $\omega_o$ itself is large. Our first objective will be to obtain this theory.

In a transverse magnetic field, the total `magnetic momenta' $P_i$ obey the nontrivial algebra $[P_i,P_j] = - i \epsilon_{ij} B N$. Here $B$ is the magnetic field and $N$ the electric charge operator. When the generators of symmetries do not commute, and have expectation values ($\langle N \rangle \neq 0$ in this case), the number of Goldstone bosons that arise can be fewer than the number of spontaneously broken symmetries \cite{PhysRevLett.108.251602}. The effective Lagrangian for the Goldstone fields can be first order in time derivatives, in such a way that the fields are not all independent degrees of freedom. In particular, 
let $\varphi_i$ be the Goldstone fields corresponding to spontaneously broken magnetic translations. As always, the symmetries act upon the Goldstone fields by shifts, so that under a translation by $\delta x_j$ one has $\varphi_j \to \varphi_j+ \delta x_j$ to leading order in fields. The most relevant term in the effective Lagrangian that is allowed by this symmetry, as well as {\sf PT} symmetry, is then $\Lag = \epsilon^{ij} \varphi_i \dot \varphi_j + \cdots$. Here $\cdots$ denotes terms that are higher order in fields (negligible in linear response hydrodynamic regimes) or spatial derivatives (to be restored shortly). Let us see how this Lagrangian encodes magnetic translations in the symmetry-broken state. Firstly, quantization immediately yields the commutator
\begin{equation}\label{eq_com_2}
[\varphi_i(x),\varphi_j(y)] = -{i\epsilon_{ij}} \delta(x-y)
\, .
\end{equation}
Using the transformation $\varphi_j \to \varphi_j+ \delta x_j$, 
the standard Noether argument implies that the conserved densities $\pi_i \propto \epsilon_{ij} \varphi^j$. Thus using \eqref{eq_com_2}, and fixing the normalization, we indeed reproduce the magnetic translation algebra with
\be\label{eq:PP}
P_i = \int \pi_i d^2x = \sqrt{n B} \int \epsilon_{ij} \varphi^j d^2x \,.
\ee
Within linearized hydrodynamics $N$ can be replaced by its expectation value, and hence $n = \langle N\rangle/\text{Vol}$ is the charge density.

Allowing for pinning --- i.e. weak explicit breaking of magnetic translations --- and restoring spatial gradient terms, the non-dissipative, long-wavelength and linearized dynamics of the pseudo-Goldstone fields is thus described by, with $k$ the wavevector,
\begin{equation}\label{eq_L}
\mathcal L
	= \epsilon^{ij} {\varphi_i}\dot\varphi_j -  \varphi_i \left[\delta^{ij}\omega_\text{pk} + \left(\kappa k^i k^j + \mu k^2\delta^{ij} \right) + \ldots\right]\varphi_j \, .
\end{equation}
We assumed isotropy ($6$-fold rotation symmetry is sufficient \cite{PhysRevB.96.195128}) and {\sf PT} symmetry to restrict the form of the spatial derivative terms. 
The pinning term $\omega_\text{pk}$ breaks the invariance under translations
$\varphi^i\to \varphi^i+\delta x^i$. The stiffnesses and pinning frequency must satisfy $\omega_\text{pk},\, \mu,\,\kappa>0$ for the potential to be positive definite. Beyond this fact, they are a priori undetermined coefficients in the derivative expansion.

The equations of motion following from \eqref{eq_L} lead to the dispersion relation
\begin{equation}\label{eq_disp}
\omega(k) = \pm \sqrt{\left(\omega_\text{pk} + \mu k^2 \right)\left(\omega_\text{pk} + (\mu+\kappa)k^2\right)} \, .
\end{equation}
In the absence of pinning we find a pair of gapless propagating modes with dispersion relation $\omega\sim\pm k^2$. These are the magnetophonons.\footnote{Coupling to three-dimensional photons changes the dispersion relation at small wavevector to $\omega\sim k^{3/2}$ \cite{FUKUYAMA19751323, PhysRevB.18.6245}. The dressing of our results by Coulomb interactions is discussed in the Appendix where, among other things, we recover this result. We will mainly be interested in the frequency-dependent conductivity, which is the response to the total (rather than external) electric field, and therefore we need only consider the unscreened response to obtain $\sigma(\omega,k)$, see e.g. \cite{COTE1992187}. The experimental detection of a screened magnetophonon has recently been reported \cite{Jang2016}.} We will mostly be interested in the limit $k \to 0$, wherein \eqref{eq_disp} gives the pinned magnetophonon gap $\omega=\pm \omega_\text{pk}$. 

Dissipation and coupling to charge fluctuations is added to the theory following the usual constitutive relations of hydrodynamics \cite{chaikin1995principles}. Setting $k=0$, these can be written usefully as
\be\label{eq:mat}
\left(
\begin{array}{c}
j^i \\
\dot \varphi^i
\end{array}
\right)
= \left(
\begin{array}{cc}
\sigma^{ij}_0 & \gamma^{ij}\\
\gamma^{ij} & \Omega^{ij}/\omega_\text{pk}
\end{array}
\right)
\left(
\begin{array}{c}
E_j \\
s_j -\omega_\text{pk} \varphi_j
\end{array}
\right) \,.
\ee
Here $E_j$ is an external electric field and $s_j$ is the source for $\varphi_j$. In equilibrium $\varphi_j = s_j/\omega_{\rm pk}$. This formula describes the electric current coupled to an additional slow mode. We will ignore the coupling to thermal currents for simplicity, this can be straightforwardly incorporated. Equality of the off-diagonal terms in the above matrix follows from Onsager reciprocity, as we show in the Appendix. The transport coefficients appearing in \eqref{eq:mat} have longitudinal and Hall components, so that $\sigma_0^{ij} = \sigma_0 \delta^{ij} + \sigma_0^H \epsilon^{ij}$, $\gamma^{ij} = \gamma \delta^{ij} + \gamma_H \epsilon^{ij}$ and $\Omega^{ij} = \Omega \delta^{ij} + \omega_\text{pk} \epsilon^{ij}$. The physical meaning of these terms will become apparent shortly. The dissipative coefficients are $\sigma_0, \gamma,\Omega$ and positivity of entropy production requires $\gamma^2 \leq \sigma_0 \Omega/\omega_\text{pk}$.

By eliminating $\varphi_i$ in \eqref{eq:mat}, setting the sources $s_i = 0$, and writing Ohm's law as $j_i = \sigma_{ij} E^j$ we obtain the frequency-dependent conductivities
\begin{subequations}\label{eq:www}
\begin{align}\label{eq:ww}
\sigma_{xx}(\omega)&=\sigma_0+ b \, \omega_\text{pk}\frac{(1-a^2)(-i\omega+\Omega)-2a\omega_\text{pk}}{(-i\omega+\Omega)^2+\omega_\text{pk}^2}\,,\\
\sigma_{xy}(\omega)&=\sigma_0^H+ b \, \omega_\text{pk} \frac{-2a(-i\omega+\Omega)+(a^2-1)\omega_\text{pk}}{(-i\omega+\Omega)^2+\omega_\text{pk}^2}\,.
\end{align}
\end{subequations}
Here we set $a\equiv\gamma/\gamma_H$ and $b \equiv \gamma_H^2$. The physical meaning of the various terms is apparent: $\sigma_0$ and $\sigma_0^H$ describe current dissipation into modes other than the magnetophonon, $\Omega$ is the phase relaxation rate, setting the width of the peak, $\omega_\text{pk}$ is the pinning frequency, $a$ determines the deviation from a strict Lorentzian form, and $b$ controls the spectral weight of the peak.

The expressions \eqref{eq:www} are similar to results obtained in microscopic theories of crystal vibrations \cite{PhysRevB.18.6245,PhysRevB.46.3920, PhysRevLett.80.3827, PhysRevB.65.035312, PhysRevB.62.7553}. Our theory has directly zoomed in on the low energy magnetophonon, whereas microscopic descriptions necessarily include the high energy magnetoplasmon also. For completeness, in the Appendix we give a hydrodynamic discussion of the full magnetophonon/magnetoplasmon system. The other difference is that because \eqref{eq:www} holds at low frequencies below the local thermalization rate $\omega \ll 1/\tau_\text{eq}$, the expressions are ratios of analytic function of frequencies. In this regime, any degrees of freedom that are gapless at zero temperature with non-trivial scaling exponents (cf. \cite{PhysRevB.62.7553}) will instead produce non-analyticities in the temperature dependence of the transport coefficients.

The linear response relation \eqref{eq:mat} also leads to Kubo formulae for the dissipative coefficients. In particular,
\begin{subequations}\label{eq_Kubo}
\begin{align}\label{eq_Kubo_Omega}
\Omega
	&= \omega_\text{pk} \lim_{\omega\to 0}\lim_{\epsilon\to 0} \frac{1}{\omega}\Im G^R_{\dot\varphi_x\dot\varphi_x}(\omega) \,, \\
\gamma\label{eq_Kubo_gamma}
	&
	= \lim_{\omega\to 0}\lim_{\epsilon\to 0} \frac{1}{\omega} \Im G^R_{j_x \dot \varphi_x}(\omega)\, , \\
\sigma_0 &
    = \lim_{\omega\to 0}\lim_{\epsilon\to 0} \frac{1}{\omega} \Im G^R_{j_x j_x}(\omega)\, .
\end{align}
\end{subequations}
The first $\epsilon \to 0$ limit is one in which the effects responsible for $\Omega$ and $\omega_\text{pk}$ (e.g. disorder) become small. Indeed, consistency with the collective description requires $\Omega$ and $\omega_\text{pk}$ to be slow compared to the local thermalization rate. The order of limits in \eqref{eq_Kubo} is important, and is explained in more detail in the Appendix. The need to set the term $\omega_\text{pk} \varphi_j$ to zero in order to obtain Kubo formulae is visible already from \eqref{eq:mat}; this is achieved by the $\epsilon \to 0$ limit.

\pagebreak

One of the nondissipative terms can be evaluated explicitly, using the following expression (this expression is plausible from \eqref{eq:mat}, see Appendix for a detailed discussion)
\be\label{eq:gammaH}
\gamma_H  = \lim_{\omega\to 0}\lim_{\epsilon\to 0} \Re G^R_{\varphi_y j_x}(\omega) = \chi_{\varphi_y j_x} \,.
\ee
This term is therefore a susceptibility in the clean $\epsilon = 0$ system. 
From \eqref{eq:PP}, $\varphi_i$ is related to the generator of magnetic translations by $P_i = \sqrt{nB} \epsilon_{ij}\varphi^{k=0}_j$. Therefore,
\be\label{eq:gammaHnu}
\gamma_H  = \frac{1}{\sqrt{n B}}\chi_{P_x j_x} = \sqrt{\nu} \,,
\ee
where in the last step $\chi_{P j} = n$ follows from a standard argument \cite{forster1975hydrodynamic}, based on the fact that $\langle [P_x, j_x] \rangle \sim \pa_x j_x \sim \dot n$. Here the filling fraction
$\nu \equiv n/B$. This leads to the advertized results \eqref{eq:sintro} and \eqref{eq:S} for the optical conductivity and spectral weight in the magnetophonon peak. In (\ref{eq:sintro}) we dropped the $\sigma_0$ term to focus on the form of the peak. See also \S \ref{sec:exp} below.

\section{Microscopic evaluation of Kubo formulae}
\label{sec:dis}

The dissipative coefficients describing phase relaxation, $\Omega$ and $\gamma$, are to be determined via the Kubo formulae \eqref{eq_Kubo}.
To use these formulae we must first obtain the operator $\dot \varphi = i [H,\varphi]$. In particular, we need the decay of the homogeneous phase mode $\varphi_i^{k=0} = \int d^2x \varphi_i(x)$. Different mechanisms are possible, corresponding to different terms in $H$ that have a nonzero commutator with $\varphi_i^{k=0}$. Here $H$ is not the microscopic Hamiltonian but should be thought of as giving the leading corrections away from the dissipationless effective theory \eqref{eq_L}. We shall describe two mechanisms, corresponding to phase relaxation due to disorder and mobile dislocations.

\subsection{Phase dissipation due to disorder}

Explicit, microscopic breaking of translational symmetry, such as by disorder, allows various new terms to appear in the effective long wavelength Hamiltonian. We have already allowed for the pinning term $\omega_\text{pk} \varphi^2$. That term alone, however, does not lead to phase relaxation, which requires dissipation of the phase into other modes. For this we must consider more general terms of the form
\be\label{eq:Hdis}
H_\text{dis} = \lambda \int d^2x \varphi_i(x) \ocal_i(x) \,.
\ee
Here $\ocal_i$ is a local (vector) operator and $\lambda$ a coupling. This Hamiltonian breaks translation invariance --- recall that $\varphi$ transforms by a shift --- thereby encoding the microscopic disorder.

Given the term \eqref{eq:Hdis} in the Hamiltonian, the commutator \eqref{eq_com_2} of the phase operator with itself leads to
\be\label{eq:opeq}
\dot \varphi_i^{k=0} = i [H_\text{dis}, \varphi_i^{k=0}] = - \lambda \epsilon_{ij} \ocal_j^{k=0} \,.
\ee
The commutator of $\varphi$ with $\ocal$ in \eqref{eq:Hdis} contributes an additional term
$\dot \varphi \sim \frac{1}{\sqrt{B}} \varphi \pa \ocal$, using \eqref{eq:PP}. This composite operator can be neglected in hydrodynamic regimes, and is also small at large fields.
The Kubo formulae \eqref{eq_Kubo} then become
\begin{align}\label{eq:univ1}
\Omega_\text{dis} & = \omega_\text{pk} \lambda^2 \lim_{\omega\to 0}\lim_{\epsilon\to 0} \frac{1}{\omega}\Im G^R_{\ocal_y\ocal_y}(\omega,k=0) \,, \\
\gamma_\text{dis} & = -\lambda \lim_{\omega\to 0}\lim_{\epsilon\to 0} \frac{1}{\omega}\Im G^R_{j_x\ocal_y}(\omega,k=0) \,. 
\end{align}

An especially universal coupling is to the current operator, so that $\ocal_i = \epsilon_{ij} j_j$. The factor of $\epsilon$ is necessary for the coupling to respect {\sf PT} symmetry. This term is in fact always present, and the coupling constant is fixed as
\be\label{eq:niceH}
H_\text{dis} = \frac{1}{\gamma^H} \int d^2x \epsilon_{ij} \varphi_i(x) j_j(x) \,.
\ee
This interaction leads to $\dot \varphi_i = j_i/\gamma^H$, which is exactly the relationship obtained by using (\ref{eq:mat}) to solve for $\dot \varphi$ in terms of $j$ in the absence of sources and dissipation. The term (\ref{eq:niceH}) is also consistent with the expression (\ref{eq:gammaH}) for $\gamma^H$ as a susceptibility.

Recalling from (\ref{eq:gammaHnu}) that $\gamma^H = \sqrt{\nu}$, the coupling (\ref{eq:niceH}) immediately leads to
\be\label{eq:univ}
\Omega_\text{dis} = \frac{\omega_\text{pk} \sigma_0}{\nu} \,, \quad \gamma_\text{dis} = \frac{\sigma_0}{\sqrt{\nu}} \,.
\ee
These expressions describe dissipation of the pinned phase into charge carriers other than the magnetophonon mode. Broken translation invariance allows the phase to mix with the current according to (\ref{eq:niceH}), and the current is then able to dissipate. This physics has a similar flavor to that explored in \cite{PhysRevB.46.3920}, but the results are not the same. In particular, the expressions in \eqref{eq:univ} saturate the entropy production bound given below \eqref{eq:mat} above: $\gamma^2 = \sigma_0 \Omega/\omega_\text{pk}$. Furthermore, the dc conductivity following from (\ref{eq:ww}) vanishes with the values (\ref{eq:univ}). These are distinctive features.\footnote{A closely related coupling is responsible for saturation of the entropy production bound at low temperatures in a (zero field) holographic model \cite{Amoretti:2018tzw}.} The contributions \eqref{eq:univ} are universally present, independently of the pinning mechanism.

If the disorder is long wavelength enough to be sensibly described within the effective Hamiltonian, one can in addition consider terms of the form $H_\text{dis} = \int d^2x V(x) \ocal(x)$, with a disorder potential $V(x)$. The corresponding phase relaxation now comes from the fact that $\varphi$ acts on $\ocal$ as a magnetic translation $P$. The resulting expressions are similar to those for the momentum relaxation rate obtained in \cite{Hartnoll:2008hs,Hartnoll:2016apf}. Indeed, writing $\varphi_j = \epsilon_{ij} \pi_i/\sqrt{nB}$, from \eqref{eq:PP}, the Kubo formula \eqref{eq_Kubo_Omega} gives $\Omega_\text{dis} = \omega_\text{pk} \Gamma/\omega_c$, with $\Gamma$ the momentum relaxation rate. These terms will typically be small in the large field limit.\footnote{Operators such as $\ocal = \varphi^2$ are exceptions, because magnetic translations act on $\varphi$ as a shift, with no factors of $1/B$. It can be verified that such terms do not contribute to phase relaxation when the disorder wavelengths are long enough to be included within hydrodynamics.}

\subsection{Phase dissipation due to mobile dislocations}

The phase can also be relaxed by mobile dislocations, which are topological defects describing a vortex in the translational order. Dislocations can relax the phase even without disorder and pinning, and may be expected to become important during the melting of the solid \cite{PhysRevLett.66.652}. To incorporate vortex dynamics it is necessary to keep track of some spatial gradients. With non-vanishing gradients, the equation of motion for the phase (in the absence of pinning or dissipation, i.e. inside the $\epsilon \to 0$ limit) is
\begin{equation}\label{eq_phidot}
\dot\varphi_i
	= i [H,\varphi_i]
	= {\epsilon_{ik} }\left(\kappa \d^k\d^j + \mu\nabla^2 \delta^{kj}\right)\varphi_j - \frac{\gamma_H}{\chi} \epsilon_{ij}\d_j n + \ldots\, ,
\end{equation}
The first terms here follow immediately from the Lagrangian \eqref{eq_L}, while the non-dissipative coupling to the charge density $n$ is determined from \eqref{eq:mat}: Gradients are restored in \eqref{eq:mat} by letting $E \to E - \nabla \mu_e = E - (\nabla n)/\chi$. Here $\mu_e$ is the chemical potential
and $\chi$ the charge compressibility. It is clear in \eqref{eq_phidot} that the phase is not relaxed as the wavevector $k \to 0$, this is the usual protection of Goldstone bosons. However, in the cores of dislocations the symmetry is restored and the phase is not well-defined. In this case, the arguments developed in \cite{PhysRevB.94.054502, PhysRevB.96.195128, PhysRevB.97.220506} show that
\begin{equation}\label{eq:BS}
\dot\varphi_i^{k = 0} =
\frac{2 \gamma_H}{\chi} \epsilon_{ij}\int_{{\rm cores}}\!\!\!\!\! d^2 x \, \d_j n\,.
\end{equation}
This expression is closely related to the final term in \eqref{eq_phidot}, but the factor of 2 is a little subtle and is due to the fact that the integral is over all {\it mobile} dislocation cores, that have time-dependent locations 
\cite{PhysRevB.97.220506}. The first terms in (\ref{eq_phidot}) do not contribute since the phase decays rapidly in the normal state. If the symmetry is only partially restored in the core, c.f.~\cite{PhysRevB.22.2514}, there can be additional contributions to (\ref{eq:BS}).
Equation \eqref{eq:BS} should be thought of as describing a generalized Bardeen-Stephen effect in which dissipation occurs in the cores of mobile vortices \cite{PhysRev.140.A1197}.

We can now use \eqref{eq:BS} in the Kubo formulae. A simple case arises when the cores are large enough that charge diffuses within the cores. The argumentation in this case is identical to that in 
\cite{PhysRevB.97.220506}, giving the Bardeen-Stephen-like expressions
\be\label{eq:disloc}
\Omega_\text{vor} =  \frac{2 x}{\sigma_\text{n}} \nu \omega_\text{pk}  \,, \quad \gamma_\text{vor} = x \sqrt{\nu} \frac{ \sigma_\text{n}^H}{\sigma_\text{n}} \,.
\ee
Here $x$ is the fraction of the total area covered by mobile vortex cores, while $\sigma_\text{n}$ and $\sigma_\text{n}^H$ are the longitudinal and Hall conductivities of the normal state in the core. We recalled that $\gamma_H^2 = \nu$. Furthermore, at low temperatures one can expect that $\sigma_\text{n}^H = \nu$, leading to the relation $\Omega_\text{vor}/\omega_\text{pk} = 2 \gamma_\text{vor}/\sqrt{\nu} = 2 a_\text{vor}$. More generally, even when the cores are not large, one still expects $\Omega \sim \gamma \sim x$.

Phase relaxation due to mobile dislocations survives in the clean limit $\omega_\text{pk} \to 0$. In this limit the factor of $\omega_\text{pk}$ in \eqref{eq:disloc} will be replaced by terms of order $\kappa/\ell_\text{vor}^2$ and $\mu/\ell_\text{vor}^2$. Here $\ell_\text{vor}$ is the radius of the vortex. This follows from the fact that the inverse phase susceptibility that appears in the Kubo formula \eqref{eq_Kubo_Omega} should, from \eqref{eq_L}, be written more generally as
$\w_\text{pk} \delta^{ij} \to \w_\text{pk} \delta^{ij} + (\k k^i k^j + \mu k^2 \delta^{ij}) + \cdots$. See \cite{PhysRevB.96.195128} for more details. 
If we write $\omega_\text{pk} = \max(\kappa,\mu)/\ell_\text{cor}^2$, for some `correlation length' $\ell_\text{cor}$, then the relative importance of pinning for dislocation-mediated phase relaxation is determined by the ratio $\ell_\text{vor}/\ell_\text{cor}$. The two regimes are physically similar.


\section{Experiments}
\label{sec:exp}

We proceed to use the formula (\ref{eq:sintro}) to fit 
the magnetophonon peaks observed undergoing a thermal melting transition in \cite{Chen2006,doi:10.1142/S0217979207042860} as well as those undergoing a quantum melting transistion in \cite{chen2005quantum}. We will find firstly that the fits are good and secondly that the dependence of the fitting parameters on temperature and filling is in good agreement with the expectations from the dissipative mechanisms discussed in the previous \S \ref{sec:dis}. We will end this section with a discussion of the validity of the collective theory for the observed peaks. Even where these peaks are unlikely to be deep in the hydrodynamic regime, the physics captured by the theory seems to be quantitatively correct. Recall that the essential simplification introduced here by hydrodynamics is to consider only a finite number of long-lived modes, the phase and the charge density, leading to the analytic-in-frequency expression (\ref{eq:sintro}).

\begin{figure}[!ht]
    \centering
    \includegraphics[height = 3.8in]{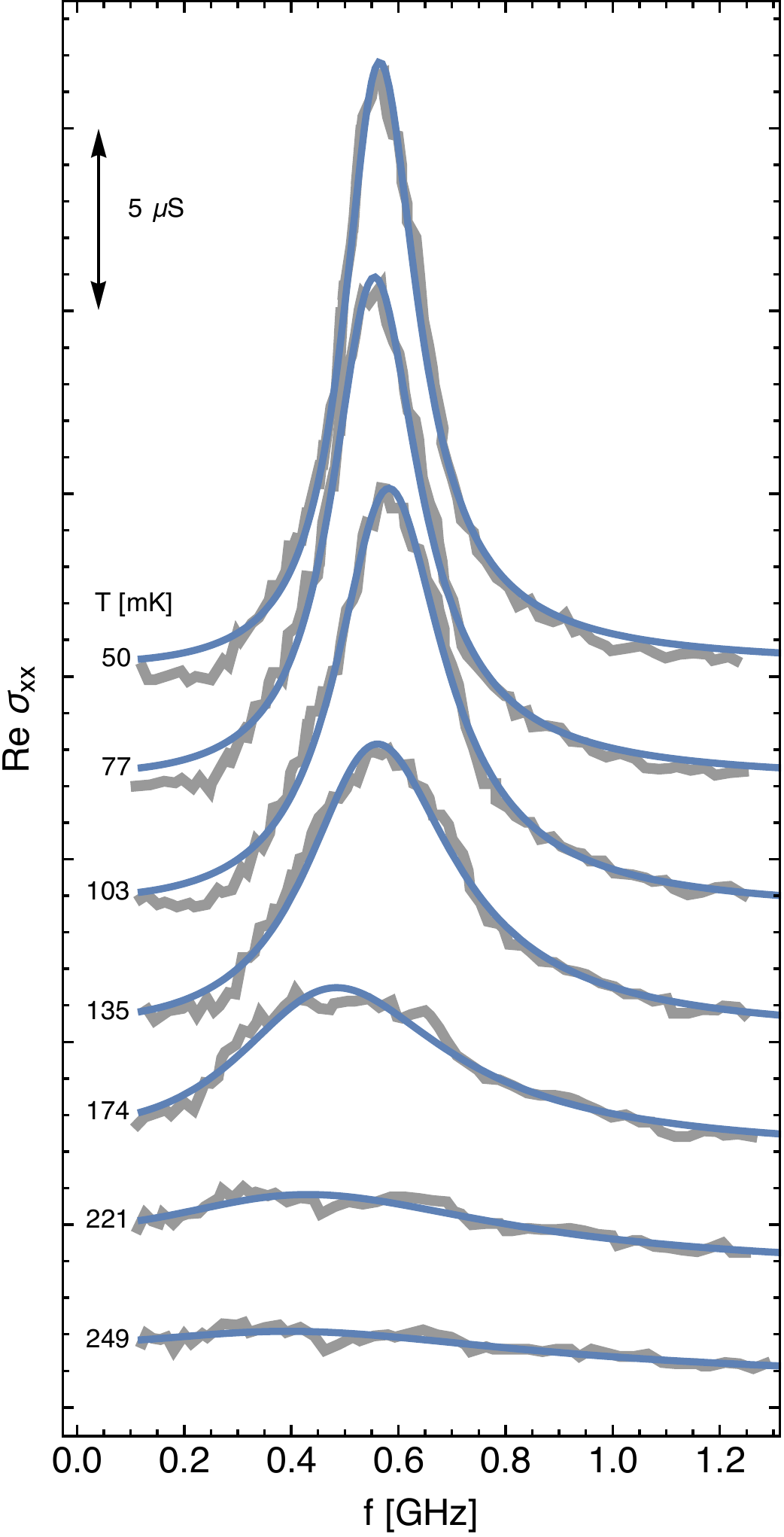}
    \includegraphics[height = 3.8in]{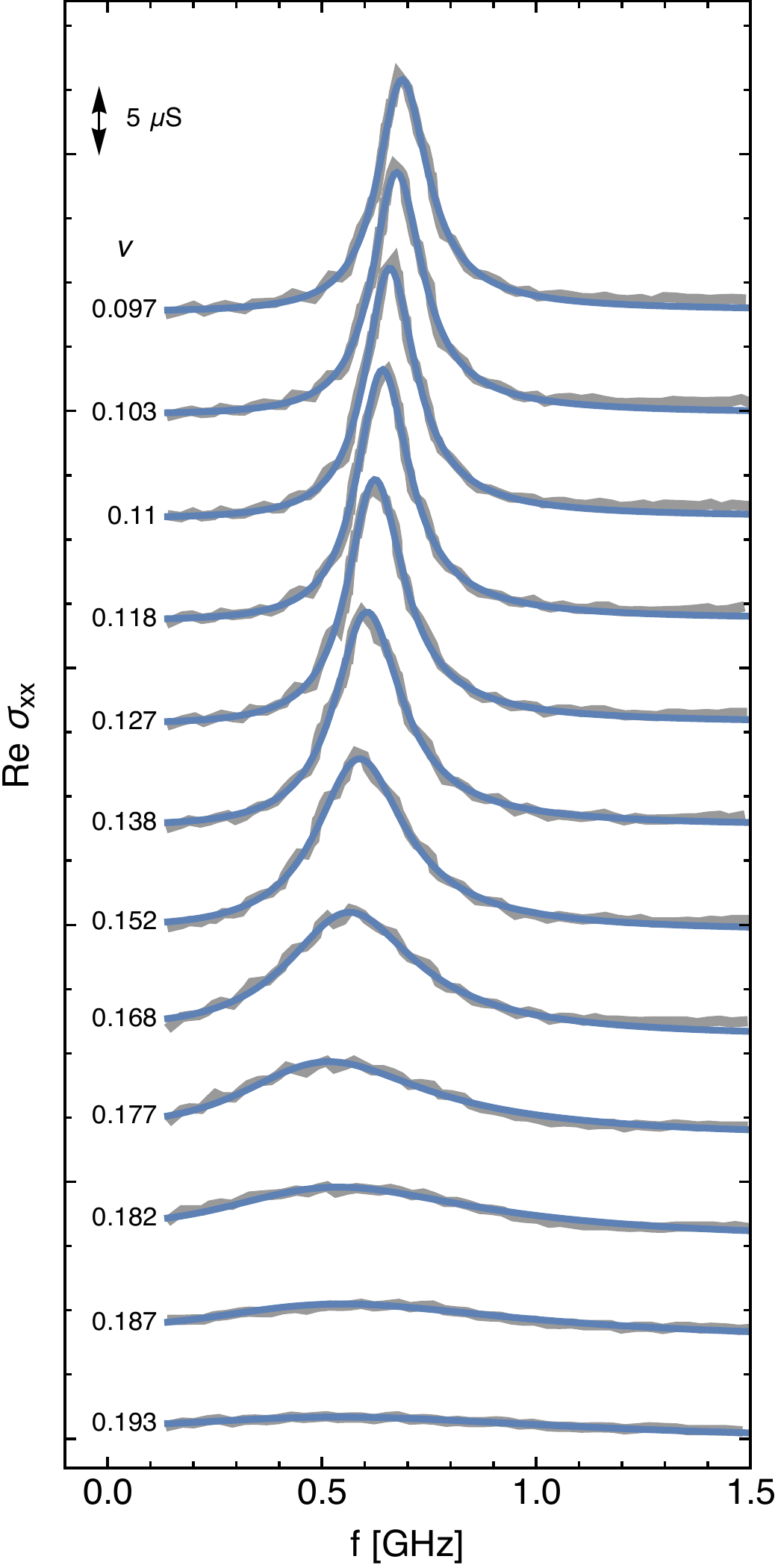}
    \includegraphics[height = 3.8in]{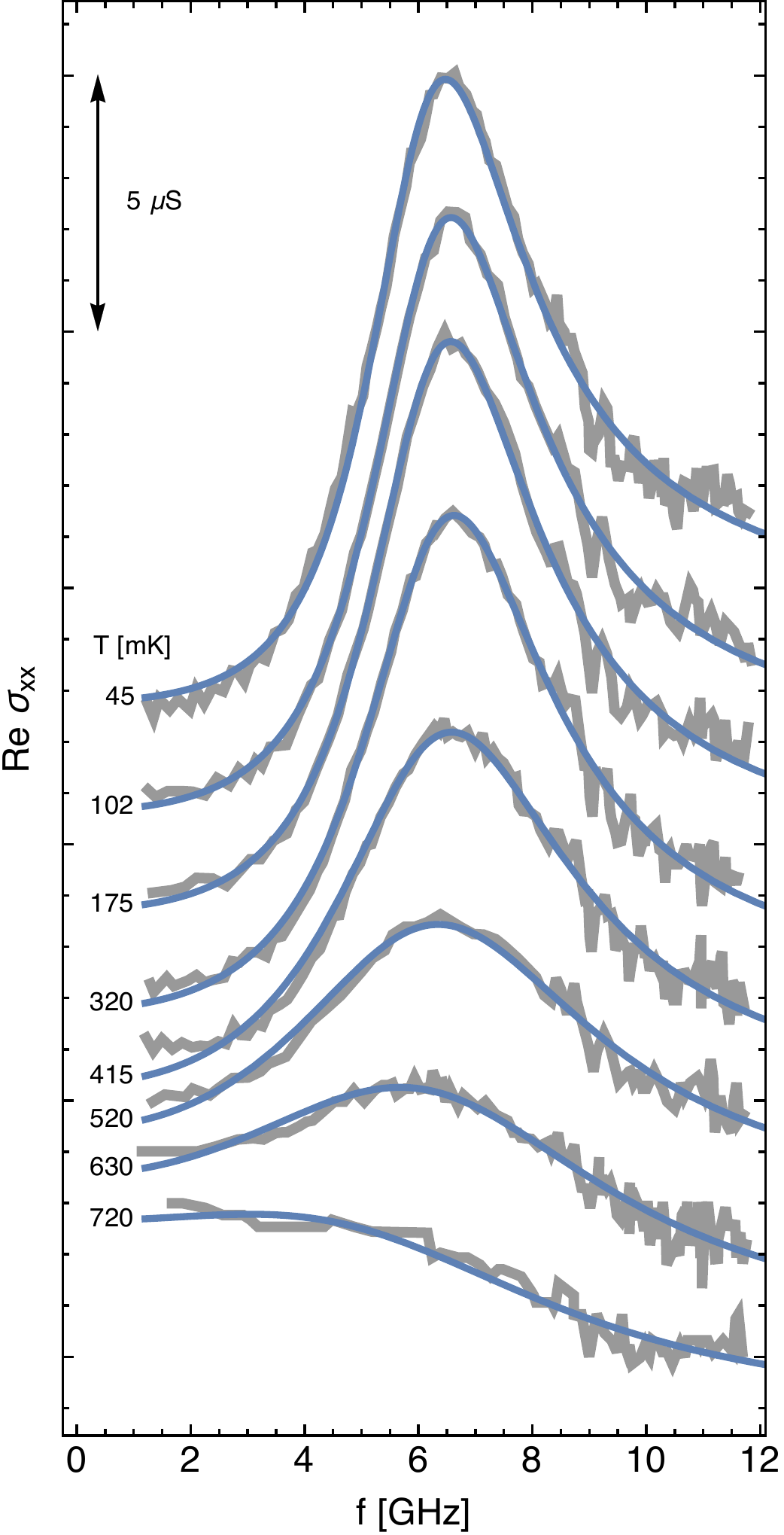}
    \caption{{\bf Fits of the magnetophonon resonance.} Blue curves are fits to (\ref{eq:sintro}). Gray is data. Curves are offset relative to each other, absolute offset has not been fit and is not shown (see main text). Left: Temperature dependence of the resonance in a sample with $\nu = 0.128$ and $B=18\, \text{T}$, data from \cite{Chen2006}. Centre: Filling fraction dependence of the resonance in a sample (`sample P') with $n=7.7\times 10^{10} \,\text{cm}^{-2}$ and $T=80\, \text{mK}$, data from \cite{chen2005quantum}.
    Right: Temperature dependence of the resonance in a sample with $\nu = 0.16$ and $B=10.3 \,\text{T}$, data from \cite{doi:10.1142/S0217979207042860}. }
    \label{fig:plots}
\end{figure}

The fits are shown in Fig. \ref{fig:plots}. In making the fits we have not attempted to fit the overall offset, controlled by $\sigma_0$ in (\ref{eq:sintro}). There are various background sources of dissipation in the experimental setup, that were addressed by subtraction of a reference function of frequency, see e.g. \cite{chen2005quantum}. This means that the shape of the curve away from the peak should perhaps not be taken too literally; for example, the data as presented included negative values of $\sigma_{xx}$ at small and large frequencies. For this reason, we have only fit the parameters $\Omega, \omega_\text{pk}, a$ that determine the location and shape of the peak. Ideally, the offset $\sigma_0$ could be found by independent measurements of the dc conductivity in the same extremely pure samples in which the peak has been measured. 

\begin{figure}[!ht]
    \centering
    \includegraphics[height = 1.9in]{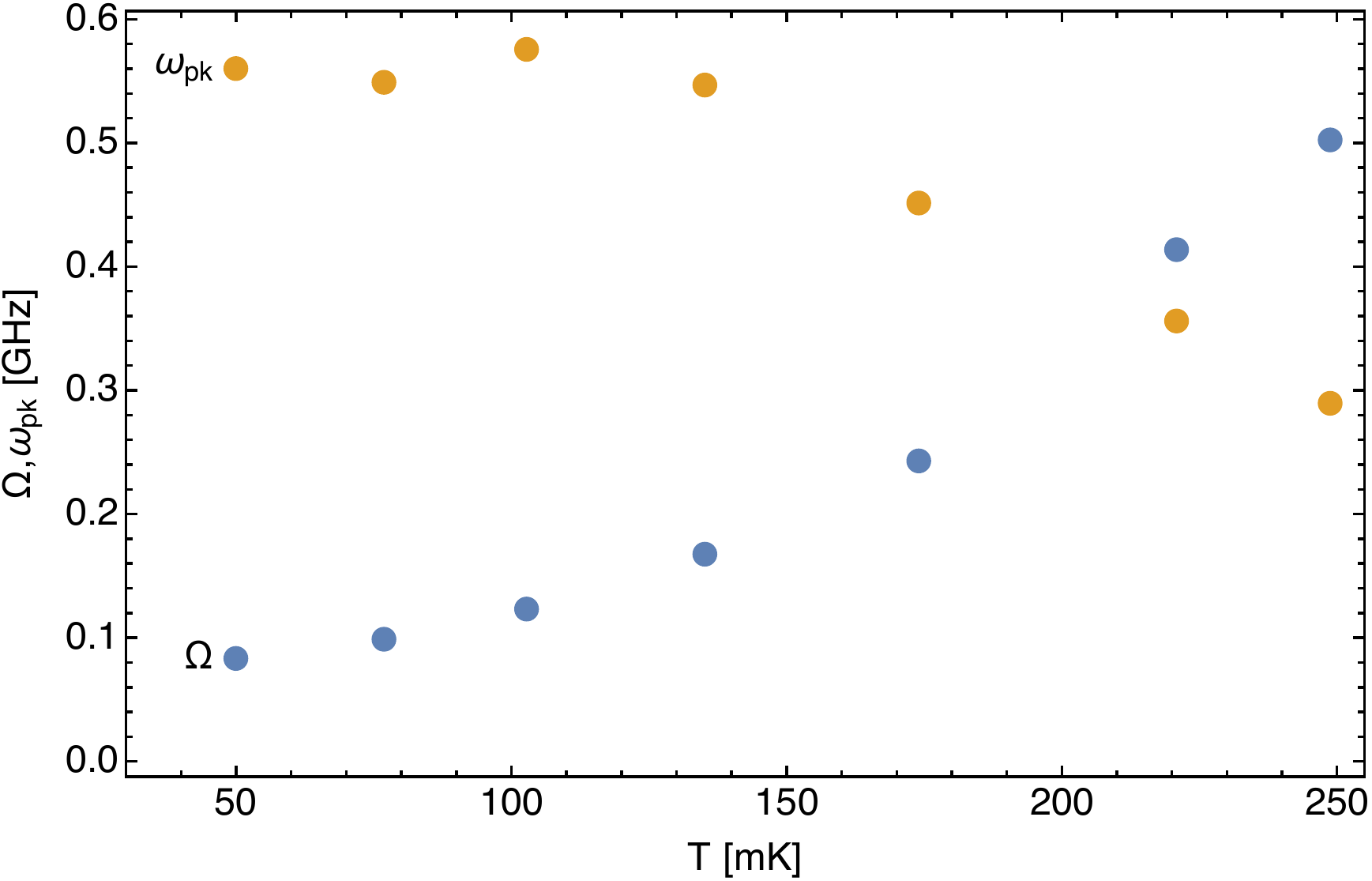}
    \includegraphics[height = 1.9in]{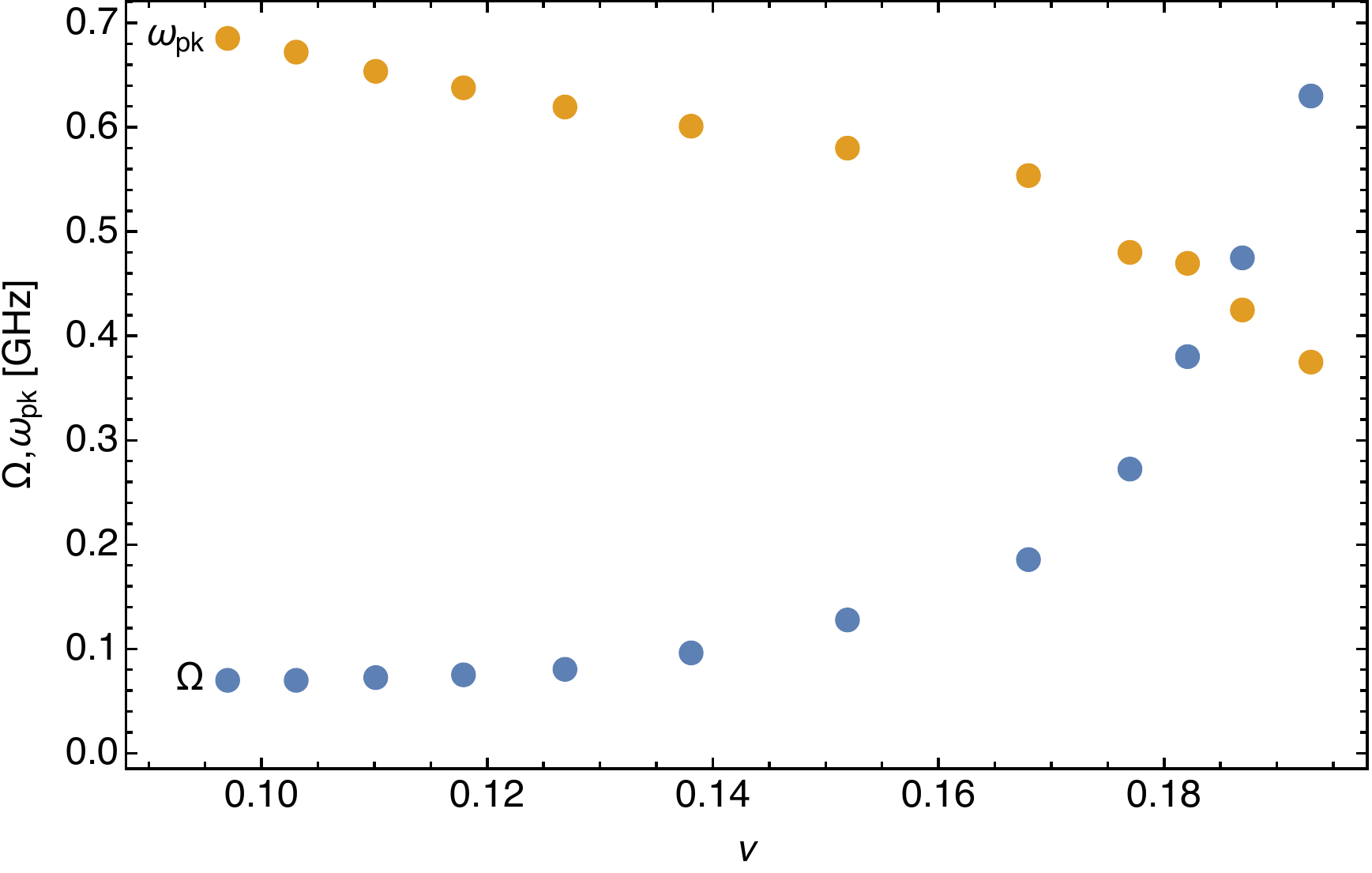}
    \caption{{\bf Thermal and quantum melting.} Variation of $\Omega$ and $\omega_\text{pk}$ as the melting transition is approached. Melting is characterized by a strong increase in $\Omega$ and a weaker decline of $\omega_\text{pk}$. Left: Thermal melting corresponding to leftmost plot in Fig. \ref{fig:plots}. Right: Quantum melting as a function of filling fraction, corresponding to the center plot in Fig. \ref{fig:plots}.}
    \label{fig:ABomegas}
\end{figure}

The plots in Fig. \ref{fig:plots} show the disappearance of the magnetophonon peak as the resonance melts. We focus first on the leftmost and center plots. These will be seen to be quantitatively very similar, despite the fact that the former shows thermal melting while the latter shows quantum melting. The fitted values of $\Omega$ and $\omega_\text{pk}$ for these two cases are shown in Fig. \ref{fig:ABomegas}. These plots show the same behavior: the width $\Omega$ increases significantly, by a factor of more than 5, while the peak frequency decreases a little, by a factor of 2. This behavior is consistent with melting by phase disordering rather than vanishing stiffness --- in the latter case $\omega_\text{pk}$ would be expected to go to zero.

\begin{figure}[!ht]
    \centering
    \includegraphics[height = 2.05in]{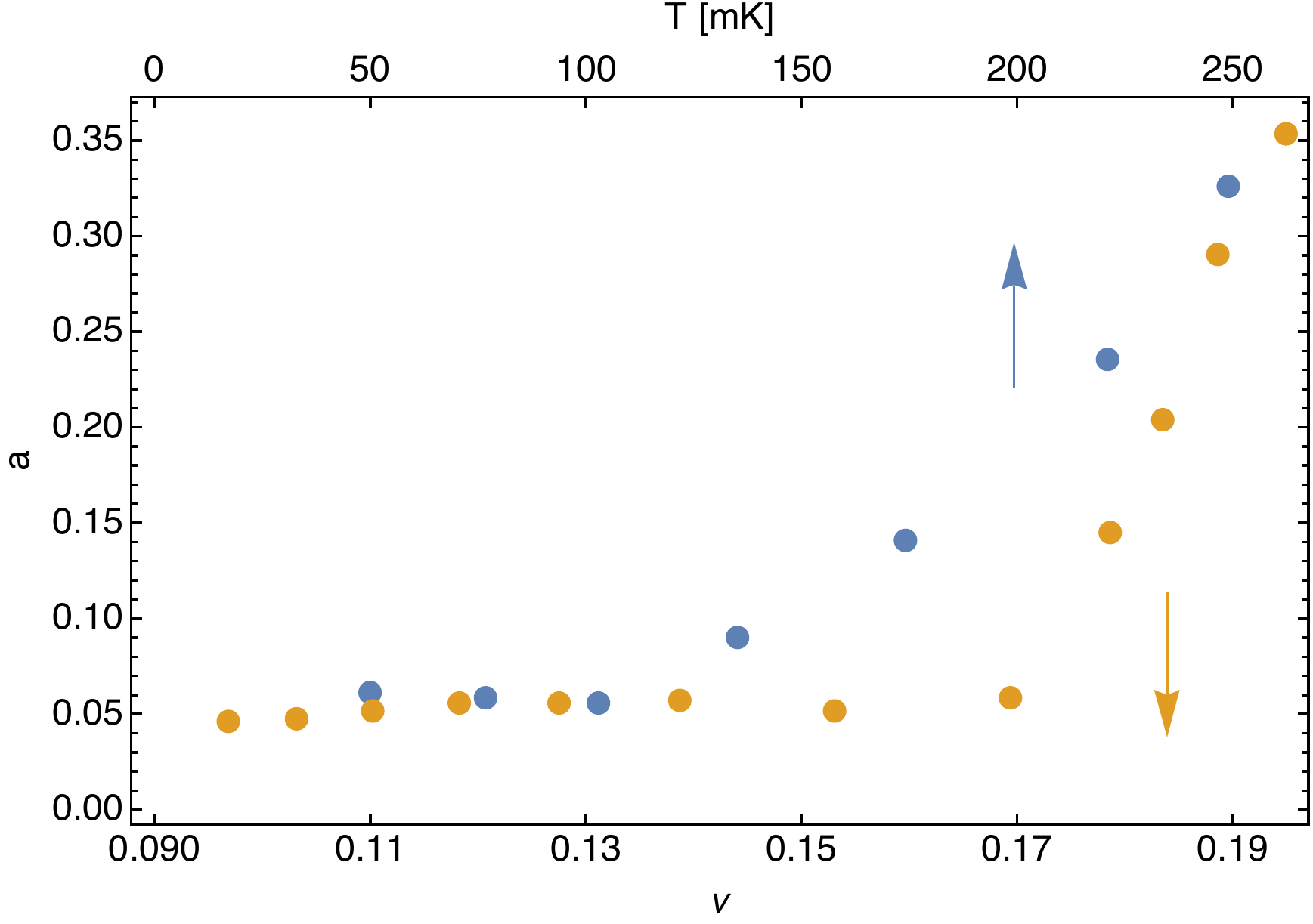}
    \includegraphics[height = 2.05in]{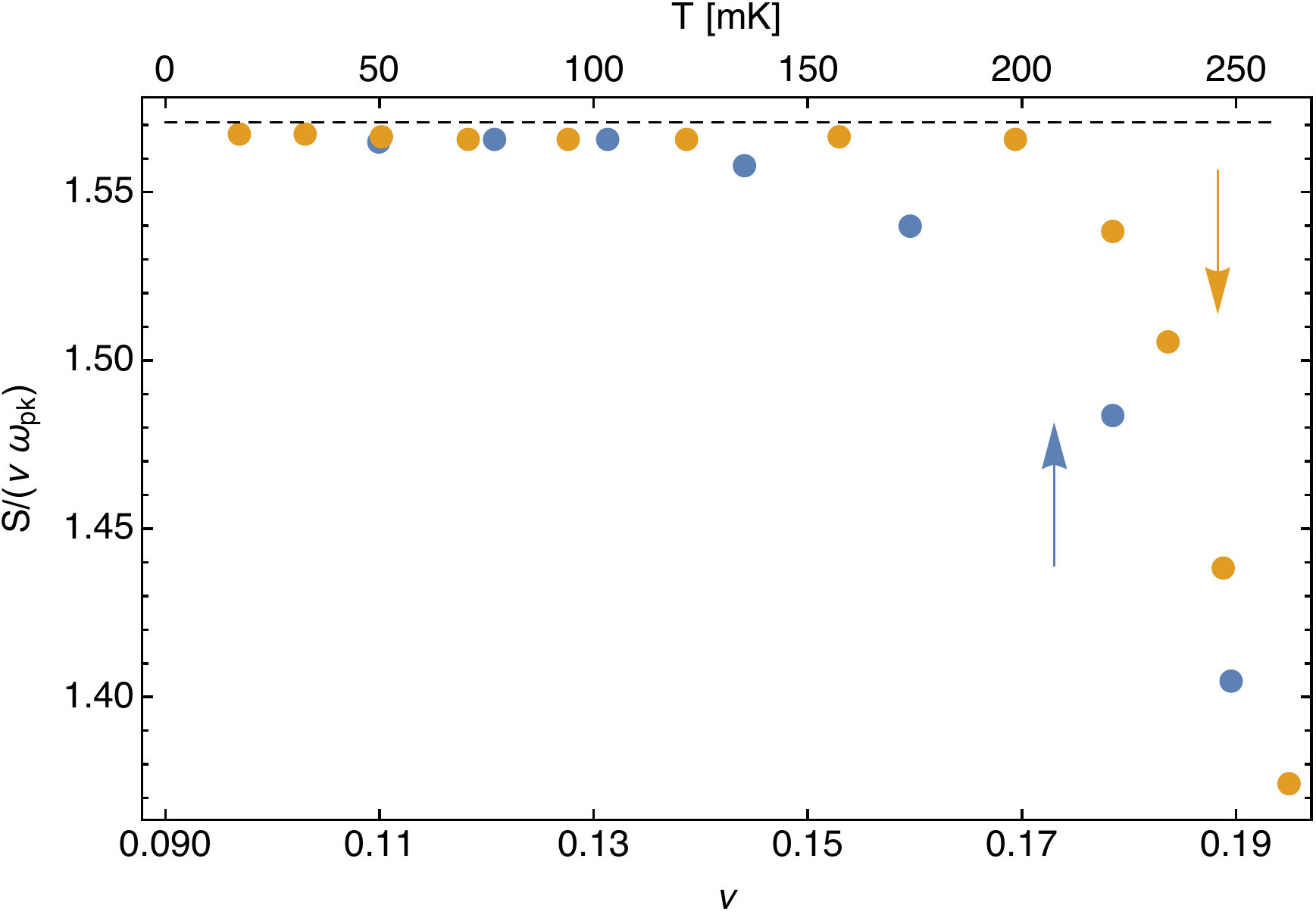}
    \caption{{\bf Non-Lorentzian peaks and violation of the Fukuyama-Lee sum rule.}  Left: The dimensionless parameter $a$, indicating departure from a purely Lorentzian peak, grows significantly as melting is approached. Shown as a function of temperature (for the leftmost plot in Fig. \ref{fig:plots}) and as a function of filling (for the centre plot in Fig. \ref{fig:plots}). Right: Corresponding violation of the Fukuyama-Lee sum rule (shown as dashed black line) for the ratio $S/(\nu \omega_\text{pk})$ as melting is approached.}
    \label{fig:FL}
\end{figure}

The `non-Lorentzian' parameter $a$ for these two sets of data is shown in Fig. \ref{fig:FL}. The left plot shows a dramatic increase in $a$ towards the melting transition, closely tracking the increase of $\Omega$ in Fig. \ref{fig:ABomegas}. The range of values taken by $a$ is the same in the thermal and quantum cases.
The right plot shows how this
increase in $a$ close to the melting transition leads to a violation of the Fukuyama-Lee sum rule, according to (\ref{eq:S}). This plot shows the spectral weight $S$ divided by $\nu \omega_\text{pk}$, with the Fukuyama-Lee result shown as a dashed black line. In addition to the cases considered here, similar departures from the Fukuyama-Lee sum rule have been widely seen in other data close to melting transitions, e.g. \cite{PhysRevLett.91.016801, SAMBANDAMURTHY2006100, PhysRevB.95.045417}.

\begin{figure}[!ht]
    \centering
    \includegraphics[height = 2.1in]{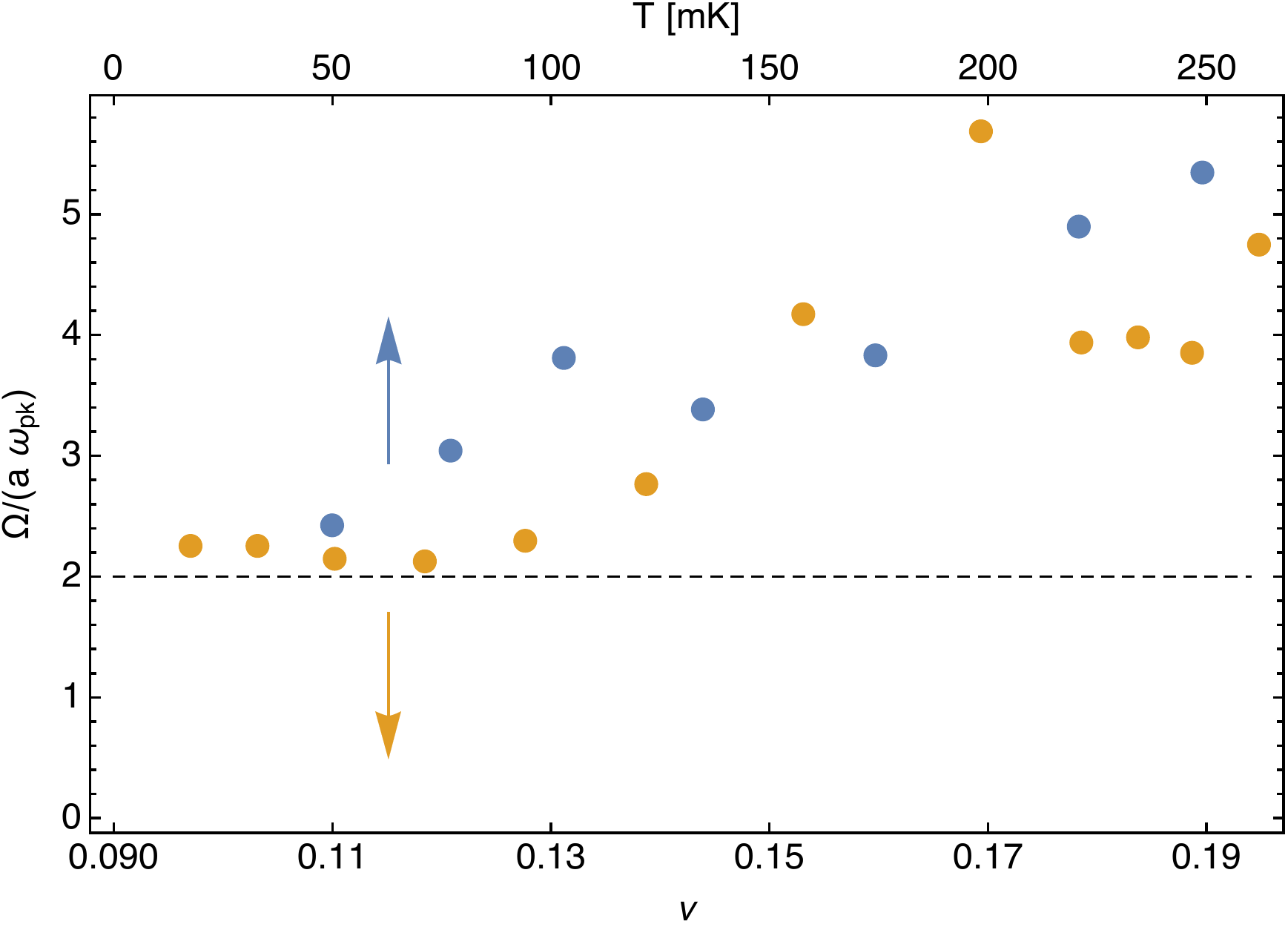}
    \includegraphics[height = 1.93in]{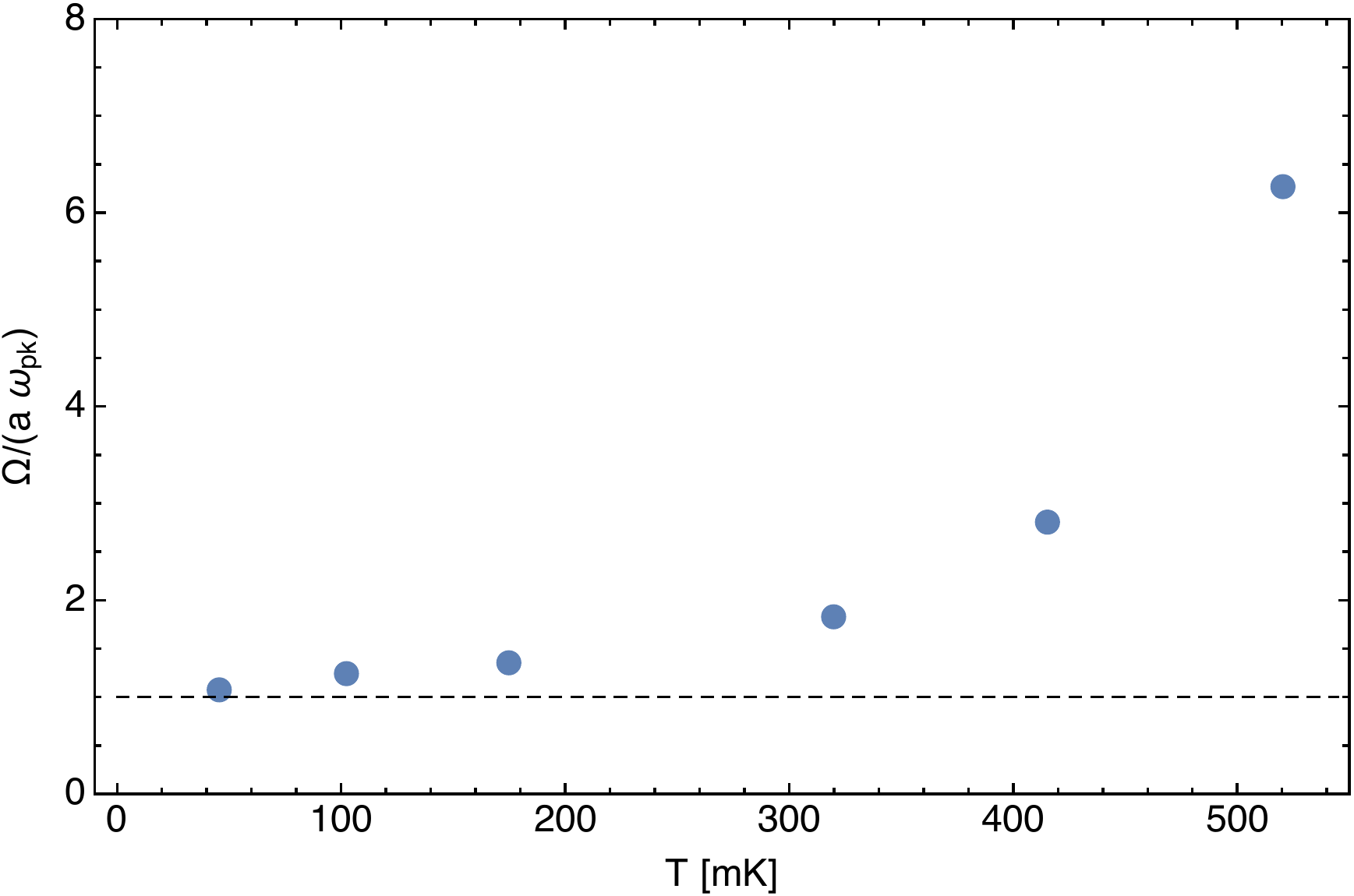}
    \caption{{\bf Ratio of coefficients suggests different dissipative mechanisms} are at work in the less disordered and more disordered samples. Left: The low temperature/low filling behavior of the cleaner samples (leftmost and center plots in Fig. \ref{fig:plots}) is consistent with dissipation into mobile dislocations, shown as a dashed black line. Right: The low temperature behavior of the more disordered sample (rightmost plot in Fig. \ref{fig:plots}) is consistent with the universal phase dissipation into currents, shown as a dashed black line.}
    \label{fig:Rat}
\end{figure}

The discussion of dissipation mechanisms in \S\ref{sec:dis} led to the expressions (\ref{eq:univ}) and (\ref{eq:disloc}) for the dissipative parameters. These in turn lead to the relations
\be\label{eq:mec}
\frac{\Omega_\text{dis}}{a_\text{dis} \, \omega_\text{pk}} = 1 \,, \qquad \frac{\Omega_\text{vor}}{a_\text{vor} \, \omega_\text{pk}} = 2 \,. 
\ee
Recall again that $a = \gamma/\gamma_H = \gamma/\sqrt{\nu}$. The first expression in (\ref{eq:mec}) describes disorder-mediated dissipation of the phase into currents, while the second describes phase dissipation due to mobile dislocations (`vortices'). In Fig. \ref{fig:Rat} we plot the ratio $\Omega/(a \, \omega_\text{pk})$ for all three families of fits in Fig. \ref{fig:plots}. Consider first the left plot. The behavior of this ratio is seen to be very similar between the thermal and quantum melting transitions.
At lower temperatures and fillings, away from the transition, the ratio in this plot approaches the value 2, associated in (\ref{eq:mec}) to mobile dislocations. This regime, with sharp peaks and hence dilute vortices --- small $x$ in (\ref{eq:disloc}) --- is precisely where the result (\ref{eq:disloc}) and hence (\ref{eq:mec}) is controlled. As the transition is approached $x \to 1$ and additional dissipative channels may also appear.

The plot of the ratio of coefficients on the right of Fig. \ref{fig:Rat} corresponds to the data on the rightmost plot of Fig. \ref{fig:plots}, that we have not discussed yet. The frequency scale on this plot in Fig. \ref{fig:plots} is an order of magnitude larger than on the other two, indicating that this is a more disordered sample (and the melting temperature is also higher). Indeed, in Fig. \ref{fig:Rat} we see that at low temperatures, the ratio now tends to the value 1, associated in (\ref{eq:mec}) to disorder rather than dislocations. As previously, as the melting temperature is approached, more dissipative channels are likely to open and operators other than current are likely to become important in (\ref{eq:Hdis}).

All told, the form (\ref{eq:sintro}) of the magnetophonon peak together with the results (\ref{eq:univ}) and (\ref{eq:disloc}) for the dissipative parameters appear to give quantitative insight into the dissipative melting dynamics of magnetophonons. Let us estimate the strength of interactions necessary for these observed peaks to be within a hydrodynamic regime.
The local thermalization time can be usefully parametrized as $\tau_\text{eq} = \a \, \hbar/(k_B T)$, with $\a \sim 1$ expected for strong interactions and $\a \gg 1$ for a weakly interacting system. For the first two plots in Fig. \ref{fig:plots}, the experiments we focussed on, the pinned magnetophonon resonances are in the range $h f_\text{pk}/(k_B T) \sim 0.1 - 0.5$. This requires $\a \lesssim 10$ for the highest temperatures and $\a \lesssim 2$ at lower temperatures. For this to hold, moderate to strong interactions are necessary. The more disordered sample, the third plot in Fig. \ref{fig:plots}, is further from the collective regime, with $h f_\text{pk}/(k_B T) \sim 0.4 - 6.4$.
Poor screening means that Coulomb interactions are potentially important in these gapped low temperature systems (cf. \cite{PhysRevLett.122.116601}). Clearly, a direct determination of the thermalization time would be desirable. At the very least, the collective approach gives a complementary perspective to the existing harmonic vibration studies, and, beyond that, seems to organize the data in an accurate and useful way.

\section{Final comments}

Our theoretical construction relies only on the symmetries of the system, either exact (charge conservation and $\sf PT$) or approximate (magnetic translations). As such it should describe other microscopic systems with the same symmetry breaking pattern. Vortex lattices in superfluids or superconductors have a similar symmetry structure. Although particle number is also spontaneously broken there, the fact that its generator can be obtained from the spontaneously broken magnetic translations $[P_i, P_j] = -i\epsilon_{ij} BN$ implies that no additional Goldstone mode is protected \cite{Watanabe:2013iia,Brauner:2014aha}. The hydrodynamics with exact magnetic translation symmetry is then identical to the theory developped here without pinning or phase relaxation, and the quadratically dispersing mode is called the Tkachenko mode in this context (see \cite{Sonin2014} for a review, and \cite{Moroz:2018noc} for a recent effective field theory approach). The analog of the magnetoplasmon is the Kohn mode, which has been automatically integrated out in our approach. Pinning and relaxation are however qualitatively different in vortex lattices, since the spontaneous breaking of particle number conservation, which remains an exact symmetry, guarantees the existence of a superfluid sound mode as $k\to 0$.
We leave the study of pinning and relaxation in these systems for future work.

Experimentally speaking, we have focussed on the well-characterized Wigner solid in GaAs/GaAlAs heterostructures. Recent results suggest that similar magnetophonon resonances can be observed in graphene in a large field \cite{Kumar2018}. Field-induced incommensurate translational order also arises in strongly correlated systems such as cuprates \cite{Wu2011,Gerber949}, and should also lead to distinctive collective modes analogous to those we have investigated here.

\section*{Acknowledgements}

We acknowledge helpful discussions with Steve Kivelson and Sergej Moroz. BG is happy to acknowledge stimulating discussions with Andrea Amoretti, Daniel Are\'an and Daniele Musso in the course of ongoing collaborations.

This work is supported by the Department of Energy, Office of Basic Energy Sciences, under Contract No. DEAC02-76SF00515. 
LVD is partially supported by the Swiss National Science
Foundation. BG is supported by the European Research Council (ERC) under the European Union’s Horizon 2020 research and innovation programme (grant agreement No 758759). BG has been supported at earlier stages of this work by the Marie Curie International Outgoing Fellowship nr 624054 within the 7th European Community Framework Programme FP7/2007-2013 and by the European Research Council (ERC) under the European Union’s Horizon 2020 research and innovation programme (grant agreements No 341222). AK is supported by the Swedish Research Council grant 2017-00328, and was at the initiation of this work supported by the Knut and Alice Wallenberg Foundation.

\bibliographystyle{ourbst}
\bibliography{CDW}

\providecommand{\href}[2]{#2}\begingroup\raggedright\begin{thebibliography}{10}

\bibitem{PhysRevB.15.1959}
L.~Bonsall and A.~A. Maradudin, {{Some static and dynamical properties of a
  two-dimensional Wigner crystal}},
  \href{http://dx.doi.org/10.1103/PhysRevB.15.1959}{Phys. Rev. B {\bf 15},
  1959, 1977}.

\bibitem{RevModPhys.60.1129}
G.~Gr\"uner, {The dynamics of charge-density waves},
  \href{http://dx.doi.org/10.1103/RevModPhys.60.1129}{Rev. Mod. Phys. {\bf 60},
  1129, 1988}.

\bibitem{PhysRevB.18.6245}
H.~Fukuyama and P.~A. Lee, {Pinning and conductivity of two-dimensional
  charge-density waves in magnetic fields},
  \href{http://dx.doi.org/10.1103/PhysRevB.18.6245}{Phys. Rev. B {\bf 18},
  6245, 1978}.

\bibitem{chaikin1995principles}
P.~M. Chaikin and T.~C. Lubensky, \emph{Principles of condensed matter
  physics}, vol.~1.
\newblock Cambridge university press Cambridge, 1995.

\bibitem{doi:10.1002/9783527617258.ch9}
M.~Shayegan, \emph{Case for the Magnetic-Field-Induced Two-Dimensional Wigner
  Crystal}, ch.~9, pp.~343--384.
\newblock John Wiley \& Sons, Ltd, 2007.

\bibitem{refId0}
J.~S\'olyom, {Wigner crystals: New realizations of an old idea},
  \href{http://dx.doi.org/10.1051/epjconf/20147801009}{EPJ Web of Conferences
  {\bf 78}, 01009, 2014}.

\bibitem{PhysRevLett.65.633}
H.~W. Jiang, R.~L. Willett, H.~L. Stormer, D.~C. Tsui, L.~N. Pfeiffer and K.~W.
  West, {Quantum liquid versus electron solid around \ensuremath{\nu}=1/5
  {L}andau-level filling},
  \href{http://dx.doi.org/10.1103/PhysRevLett.65.633}{Phys. Rev. Lett. {\bf
  65}, 633, 1990}.

\bibitem{PhysRevB.44.8107}
H.~W. Jiang, H.~L. Stormer, D.~C. Tsui, L.~N. Pfeiffer and K.~W. West,
  {Magnetotransport studies of the insulating phase around \ensuremath{\nu}=1/5
  {L}andau-level filling},
  \href{http://dx.doi.org/10.1103/PhysRevB.44.8107}{Phys. Rev. B {\bf 44},
  8107, 1991}.

\bibitem{PhysRevLett.65.2189}
V.~J. Goldman, M.~Santos, M.~Shayegan and J.~E. Cunningham, {Evidence for
  two-dimentional quantum {W}igner crystal},
  \href{http://dx.doi.org/10.1103/PhysRevLett.65.2189}{Phys. Rev. Lett. {\bf
  65}, 2189, 1990}.

\bibitem{PhysRevLett.66.3285}
F.~I.~B. Williams, P.~A. Wright, R.~G. Clark, E.~Y. Andrei, G.~Deville, D.~C.
  Glattli, O.~Probst, B.~Etienne, C.~Dorin, C.~T. Foxon and J.~J. Harris,
  {Conduction threshold and pinning frequency of magnetically induced {W}igner
  solid}, \href{http://dx.doi.org/10.1103/PhysRevLett.66.3285}{Phys. Rev. Lett.
  {\bf 66}, 3285, 1991}.

\bibitem{PhysRevLett.60.2765}
E.~Y. Andrei, G.~Deville, D.~C. Glattli, F.~I.~B. Williams, E.~Paris and
  B.~Etienne, {Observation of a magnetically induced {W}igner solid},
  \href{http://dx.doi.org/10.1103/PhysRevLett.60.2765}{Phys. Rev. Lett. {\bf
  60}, 2765, 1988}.

\bibitem{PhysRevB.45.11342}
M.~A. Paalanen, R.~L. Willett, P.~B. Littlewood, R.~R. Ruel, K.~W. West, L.~N.
  Pfeiffer and D.~J. Bishop, {rf conductivity of a two-dimensional electron
  system at small {L}andau-level filling factors},
  \href{http://dx.doi.org/10.1103/PhysRevB.45.11342}{Phys. Rev. B {\bf 45},
  11342, 1992}.

\bibitem{PhysRevB.45.13784}
M.~A. Paalanen, R.~L. Willett, R.~R. Ruel, P.~B. Littlewood, K.~W. West and
  L.~N. Pfeiffer, {Electrical conductivity and {W}igner crystallization},
  \href{http://dx.doi.org/10.1103/PhysRevB.45.13784}{Phys. Rev. B {\bf 45},
  13784, 1992}.

\bibitem{PhysRevLett.79.1353}
C.-C. Li, L.~W. Engel, D.~Shahar, D.~C. Tsui and M.~Shayegan, {Microwave
  conductivity resonance of two-dimensional hole system},
  \href{http://dx.doi.org/10.1103/PhysRevLett.79.1353}{Phys. Rev. Lett. {\bf
  79}, 1353, 1997}.

\bibitem{ENGEL1997167}
L.~Engel, C.-C. Li, D.~Shahar, D.~Tsui and M.~Shayegan, {Microwave resonances
  in low-filling insulating phase of two-dimensional electron system},
  \href{http://dx.doi.org/10.1016/S0038-1098(97)00302-5}{Solid State
  Communications {\bf 104}, 167, 1997}.

\bibitem{ENGEL1997111}
L.~Engel, C.-C. Li, D.~Shahar, D.~Tsui and M.~Shayegan, {Microwave resonances
  in low-filling insulating phases of two-dimensional electron and hole
  systems}, \href{http://dx.doi.org/10.1016/S1386-9477(97)00025-8}{Physica E:
  Low-dimensional Systems and Nanostructures {\bf 1}, 111, 1997}.

\bibitem{HENNIGAN199853}
P.~Hennigan, A.~Beya, C.~Mellor, R.~Ga\'al, F.~Williams and M.~Henini,
  {Microwave absorption in the magnetically-induced {W}igner solid phase of a
  two-dimensional hole system},
  \href{http://dx.doi.org/10.1016/S0921-4526(98)00065-9}{Physica B: Condensed
  Matter {\bf 249-251}, 53, 1998}.

\bibitem{PhysRevB.61.10905}
C.-C. Li, J.~Yoon, L.~W. Engel, D.~Shahar, D.~C. Tsui and M.~Shayegan,
  {Microwave resonance and weak pinning in two-dimensional hole systems at high
  magnetic fields}, \href{http://dx.doi.org/10.1103/PhysRevB.61.10905}{Phys.
  Rev. B {\bf 61}, 10905, 2000}.

\bibitem{PhysRevLett.89.176802}
P.~D. Ye, L.~W. Engel, D.~C. Tsui, R.~M. Lewis, L.~N. Pfeiffer and K.~West,
  {{Correlation Lengths of the Wigner-Crystal Order in a Two-Dimensional
  Electron System at High Magnetic Fields}},
  \href{http://dx.doi.org/10.1103/PhysRevLett.89.176802}{Phys. Rev. Lett. {\bf
  89}, 176802, 2002}.

\bibitem{PhysRevLett.93.206805}
Y.~P. Chen, R.~M. Lewis, L.~W. Engel, D.~C. Tsui, P.~D. Ye, Z.~H. Wang, L.~N.
  Pfeiffer and K.~W. West, {{Evidence for Two Different Solid Phases of
  Two-Dimensional Electrons in High Magnetic Fields}},
  \href{http://dx.doi.org/10.1103/PhysRevLett.93.206805}{Phys. Rev. Lett. {\bf
  93}, 206805, 2004}.

\bibitem{Chen2006}
Y.~P. Chen, G.~Sambandamurthy, Z.~H. Wang, R.~M. Lewis, L.~W. Engel, D.~C.
  Tsui, P.~D. Ye, L.~N. Pfeiffer and K.~W. West, {{Melting of a 2D quantum
  electron solid in high magnetic field}},
  \href{http://dx.doi.org/10.1038/nphys322}{Nature Physics {\bf 2}, 452, 2006}.

\bibitem{doi:10.1142/S0217979207042860}
Y.~P. Chen, G.~Sambandamurthy, L.~W. Engel, D.~C. Tsui, L.~N. Pfeiffer and
  K.~W. West, {Microwave resonance study of melting in high magnetic field
  {W}igner solid},
  \href{http://dx.doi.org/10.1142/S0217979207042860}{International Journal of
  Modern Physics B {\bf 21}, 1379, 2007}.

\bibitem{PhysRevB.89.075310}
B.-H. Moon, L.~W. Engel, D.~C. Tsui, L.~N. Pfeiffer and K.~W. West, {{Pinning
  modes of high-magnetic-field Wigner solids with controlled alloy disorder}},
  \href{http://dx.doi.org/10.1103/PhysRevB.89.075310}{Phys. Rev. B {\bf 89},
  075310, 2014}.

\bibitem{PhysRevB.92.035121}
B.-H. Moon, L.~W. Engel, D.~C. Tsui, L.~N. Pfeiffer and K.~W. West, {{Microwave
  pinning modes near Landau filling $\ensuremath{\nu}=1$ in two-dimensional
  electron systems with alloy disorder}},
  \href{http://dx.doi.org/10.1103/PhysRevB.92.035121}{Phys. Rev. B {\bf 92},
  035121, 2015}.

\bibitem{PhysRevLett.91.016801}
Y.~Chen, R.~M. Lewis, L.~W. Engel, D.~C. Tsui, P.~D. Ye, L.~N. Pfeiffer and
  K.~W. West, {{Microwave Resonance of the 2D Wigner Crystal around Integer
  Landau Fillings}},
  \href{http://dx.doi.org/10.1103/PhysRevLett.91.016801}{Phys. Rev. Lett. {\bf
  91}, 016801, 2003}.

\bibitem{PhysRevLett.105.126803}
H.~Zhu, Y.~P. Chen, P.~Jiang, L.~W. Engel, D.~C. Tsui, L.~N. Pfeiffer and K.~W.
  West, {{Observation of a Pinning Mode in a Wigner Solid with
  $\ensuremath{\nu}=1/3$ Fractional Quantum Hall Excitations}},
  \href{http://dx.doi.org/10.1103/PhysRevLett.105.126803}{Phys. Rev. Lett. {\bf
  105}, 126803, 2010}.

\bibitem{PhysRevB.95.045417}
A.~T. Hatke, Y.~Liu, L.~W. Engel, L.~N. Pfeiffer, K.~W. West, K.~W. Baldwin and
  M.~Shayegan, {{Microwave spectroscopic observation of a Wigner solid within
  the $\ensuremath{\nu}=1/2$ fractional quantum Hall effect}},
  \href{http://dx.doi.org/10.1103/PhysRevB.95.045417}{Phys. Rev. B {\bf 95},
  045417, 2017}.

\bibitem{PhysRevB.46.3920}
B.~G.~A. Normand, P.~B. Littlewood and A.~J. Millis, {Pinning and conductivity
  of a two-dimensional charge-density wave in a strong magnetic field},
  \href{http://dx.doi.org/10.1103/PhysRevB.46.3920}{Phys. Rev. B {\bf 46},
  3920, 1992}.

\bibitem{PhysRevB.59.2120}
H.~A. Fertig, {{Electromagnetic response of a pinned Wigner crystal}},
  \href{http://dx.doi.org/10.1103/PhysRevB.59.2120}{Phys. Rev. B {\bf 59},
  2120, 1999}.

\bibitem{PhysRevLett.80.3827}
R.~Chitra, T.~Giamarchi and P.~Le~Doussal, {{Dynamical Properties of the Pinned
  Wigner Crystal}}, \href{http://dx.doi.org/10.1103/PhysRevLett.80.3827}{Phys.
  Rev. Lett. {\bf 80}, 3827, 1998}.

\bibitem{PhysRevB.65.035312}
R.~Chitra, T.~Giamarchi and P.~Le~Doussal, {{Pinned Wigner crystals}},
  \href{http://dx.doi.org/10.1103/PhysRevB.65.035312}{Phys. Rev. B {\bf 65},
  035312, 2001}.

\bibitem{PhysRevB.62.7553}
M.~M. Fogler and D.~A. Huse, {{Dynamical response of a pinned two-dimensional
  Wigner crystal}}, \href{http://dx.doi.org/10.1103/PhysRevB.62.7553}{Phys.
  Rev. B {\bf 62}, 7553, 2000}.

\bibitem{chen2005quantum}
Y.~P. Chen, \emph{Quantum solids of two dimensional electrons in magnetic
  fields}.
\newblock PhD thesis, Princeton U, Dept. of Electrical Engineering, 2005.
\newblock
  \href{https://search.proquest.com/docview/305420029}{https://search.proquest.com/docview/305420029}.

\bibitem{PhysRevLett.66.652}
S.~T. Chui and K.~Esfarjani, {{Finite-temperature transport of a pinned 2D
  electron lattice}}, \href{http://dx.doi.org/10.1103/PhysRevLett.66.652}{Phys.
  Rev. Lett. {\bf 66}, 652, 1991}.

\bibitem{PhysRevLett.108.251602}
H.~Watanabe and H.~Murayama, {{Unified Description of Nambu-Goldstone Bosons
  without Lorentz Invariance}},
  \href{http://dx.doi.org/10.1103/PhysRevLett.108.251602}{Phys. Rev. Lett. {\bf
  108}, 251602, 2012}.

\bibitem{PhysRevB.96.195128}
L.~V. Delacr\'etaz, B.~Gout\'eraux, S.~A. Hartnoll and A.~Karlsson, {Theory of
  hydrodynamic transport in fluctuating electronic charge density wave states},
  \href{http://dx.doi.org/10.1103/PhysRevB.96.195128}{Phys. Rev. B {\bf 96},
  195128, 2017}.

\bibitem{FUKUYAMA19751323}
H.~Fukuyama, {Two-dimensional wigner crystal under magnetic field},
  \href{http://dx.doi.org/10.1016/0038-1098(75)90696-1}{Solid State
  Communications {\bf 17}, 1323, 1975}.

\bibitem{COTE1992187}
R.~C\^ot\'e and A.~MacDonald, {Frequency-dependent conductivity of a pinned
  {W}igner crystal},
  \href{http://dx.doi.org/10.1016/0039-6028(92)90334-3}{Surface Science {\bf
  263}, 187, 1992}.

\bibitem{Jang2016}
J.~Jang, B.~M. Hunt, L.~N. Pfeiffer, K.~W. West and R.~C. Ashoori, {{Sharp
  tunnelling resonance from the vibrations of an electronic Wigner crystal}},
  \href{http://dx.doi.org/10.1038/nphys3979}{Nature Physics {\bf 13}, 340,
  2016}.

\bibitem{forster1975hydrodynamic}
D.~Forster, {Hydrodynamic fluctuations, broken symmetry, and correlation
  functions},  in \emph{Reading, Mass., WA Benjamin, Inc.(Frontiers in Physics.
  Volume 47), 1975. 343 p.}, vol.~47, 1975.

\bibitem{Amoretti:2018tzw}
A.~Amoretti, D.~Are\'an, B.~Gout\'eraux and D.~Musso, {{A holographic strange
  metal with slowly fluctuating translational order}},  2018,
  [\href{http://arxiv.org/abs/arXiv:1812.08118}{{arXiv:1812.08118 [hep-th]}}].

\bibitem{Hartnoll:2008hs}
S.~A. Hartnoll and C.~P. Herzog, {{Impure AdS/CFT correspondence}},
  \href{http://dx.doi.org/10.1103/PhysRevD.77.106009}{Phys. Rev. {\bf D77},
  106009, 2008}, [\href{http://arxiv.org/abs/arXiv:0801.1693}{{arXiv:0801.1693
  [hep-th]}}].

\bibitem{Hartnoll:2016apf}
S.~A. Hartnoll, A.~Lucas and S.~Sachdev, {{Holographic quantum matter}},  2016,
  [\href{http://arxiv.org/abs/arXiv:1612.07324}{{arXiv:1612.07324 [hep-th]}}].

\bibitem{PhysRevB.94.054502}
R.~A. Davison, L.~V. Delacr\'etaz, B.~Gout\'eraux and S.~A. Hartnoll,
  {Hydrodynamic theory of quantum fluctuating superconductivity},
  \href{http://dx.doi.org/10.1103/PhysRevB.94.054502}{Phys. Rev. B {\bf 94},
  054502, 2016}.

\bibitem{PhysRevB.97.220506}
L.~V. Delacr\'etaz and S.~A. Hartnoll, {{Theory of the supercyclotron resonance
  and Hall response in anomalous two-dimensional metals}},
  \href{http://dx.doi.org/10.1103/PhysRevB.97.220506}{Phys. Rev. B {\bf 97},
  220506, 2018}.

\bibitem{PhysRevB.22.2514}
A.~Zippelius, B.~I. Halperin and D.~R. Nelson, {Dynamics of two-dimensional
  melting}, \href{http://dx.doi.org/10.1103/PhysRevB.22.2514}{Phys. Rev. B {\bf
  22}, 2514, 1980}.

\bibitem{PhysRev.140.A1197}
J.~Bardeen and M.~J. Stephen, {{Theory of the Motion of Vortices in
  Superconductors}}, \href{http://dx.doi.org/10.1103/PhysRev.140.A1197}{Phys.
  Rev. {\bf 140}, A1197, 1965}.

\bibitem{SAMBANDAMURTHY2006100}
G.~Sambandamurthy, Z.~Wang, R.~Lewis, Y.~P. Chen, L.~Engel, D.~Tsui,
  L.~Pfeiffer and K.~West, {Pinning mode resonances of new phases of 2{D}
  electron systems in high magnetic fields},
  \href{http://dx.doi.org/10.1016/j.ssc.2006.05.053}{Solid State Communications
  {\bf 140}, 100, 2006}.

\bibitem{PhysRevLett.122.116601}
H.~Deng, L.~N. Pfeiffer, K.~W. West, K.~W. Baldwin, L.~W. Engel and
  M.~Shayegan, {{Probing the Melting of a Two-Dimensional Quantum Wigner
  Crystal via its Screening Efficiency}},
  \href{http://dx.doi.org/10.1103/PhysRevLett.122.116601}{Phys. Rev. Lett. {\bf
  122}, 116601, 2019}.

\bibitem{Watanabe:2013iia}
H.~Watanabe and H.~Murayama, {{Redundancies in Nambu-Goldstone Bosons}},
  \href{http://dx.doi.org/10.1103/PhysRevLett.110.181601}{Phys. Rev. Lett. {\bf
  110}, 181601, 2013},
  [\href{http://arxiv.org/abs/arXiv:1302.4800}{{arXiv:1302.4800
  [cond-mat.other]}}].

\bibitem{Brauner:2014aha}
T.~Brauner and H.~Watanabe, {{Spontaneous breaking of spacetime symmetries and
  the inverse Higgs effect}},
  \href{http://dx.doi.org/10.1103/PhysRevD.89.085004}{Phys. Rev. {\bf D89},
  085004, 2014}, [\href{http://arxiv.org/abs/arXiv:1401.5596}{{arXiv:1401.5596
  [hep-ph]}}].

\bibitem{Sonin2014}
E.~B. Sonin, {Tkachenko waves},
  \href{http://dx.doi.org/10.1134/S0021364013240181}{JETP Letters {\bf 98},
  758, 2014}.

\bibitem{Moroz:2018noc}
S.~Moroz, C.~Hoyos, C.~Benzoni and D.~T. Son, {{Effective field theory of a
  vortex lattice in a bosonic superfluid}},
  \href{http://dx.doi.org/10.21468/SciPostPhys.5.4.039}{SciPost Phys. {\bf 5},
  039, 2018}, [\href{http://arxiv.org/abs/arXiv:1803.10934}{{arXiv:1803.10934
  [cond-mat.quant-gas]}}].

\bibitem{Kumar2018}
M.~Kumar, A.~Laitinen and P.~Hakonen, {{Unconventional fractional quantum Hall
  states and Wigner crystallization in suspended Corbino graphene}},
  \href{http://dx.doi.org/10.1038/s41467-018-05094-8}{Nature Communications
  {\bf 9}, 2776, 2018}.

\bibitem{Wu2011}
T.~Wu, H.~Mayaffre, S.~Kr{\"a}mer, M.~Horvati\'c, C.~Berthier, W.~N. Hardy,
  R.~Liang, D.~A. Bonn and M.-H. Julien, {{Magnetic-field-induced charge-stripe
  order in the high-temperature superconductor YBa$_2$Cu$_3$O$_y$}},
  \href{http://dx.doi.org/10.1038/nature10345}{Nature {\bf 477}, 191, 2011}.

\bibitem{Gerber949}
S.~Gerber, H.~Jang, H.~Nojiri, S.~Matsuzawa, H.~Yasumura, D.~A. Bonn, R.~Liang,
  W.~N. Hardy, Z.~Islam, A.~Mehta, S.~Song, M.~Sikorski, D.~Stefanescu et~al.,
  {{Three-dimensional charge density wave order in YBa$_2$Cu$_3$O$_{6.67}$ at
  high magnetic fields}},
  \href{http://dx.doi.org/10.1126/science.aac6257}{Science {\bf 350}, 949,
  2015}.

\bibitem{BEEKMAN20171}
A.~J. Beekman, J.~Nissinen, K.~Wu, K.~Liu, R.-J. Slager, Z.~Nussinov,
  V.~Cvetkovic and J.~Zaanen, {Dual gauge field theory of quantum liquid
  crystals in two dimensions},
  \href{http://dx.doi.org/10.1016/j.physrep.2017.03.004}{Physics Reports {\bf
  683}, 1, 2017}.

\end{thebibliography}\endgroup

\appendix

\section{Hydrodynamics of pinned magnetophonons}\label{app_consti}

Here we will derive the hydrodynamic Green's functions and Kubo formulae. The most general `Josephson relation' and constitutive relation for the current, at wavevector $k=0$, read
\begin{subequations}\label{seq_constirel}
\begin{align}
\dot \varphi^i
	&= \frac{\Omega^{ij}}{\omega_{\rm pk}} (s_j - \omega_{\rm pk} \varphi_j) + {\gamma}^{ij}E_j \,,   \\
j^i
	&= \widetilde\gamma^{ij}(s_j - \omega_{\rm pk}\varphi_j) + \sigma_0^{ij}E_j \,,
\end{align}
\end{subequations}
At nonzero $k$ there will be a series of corrections to the right hand side involving terms such as $\nabla \nabla \varphi$ and $\nabla n$. The magnetophonon field $\varphi_i$ and its source $s_i$ enter in the Hamiltonian corresponding to \eqref{eq_L} as
\begin{equation}\label{eq_H0}
H = \int d^2 x \,\omega_{\rm pk}\varphi_i\varphi_i - s_i \varphi_i + \cdots\, . 
\end{equation}
The field and source must therefore appear in the combination $\omega_{\rm pk} \varphi_i - s_i$ in \eqref{seq_constirel}, in order for the current and $\dot\varphi$ to vanish at equilibrium, wherein $E=0$ and $\varphi_i = s_i/\omega_{\rm pk}$. Isotropy requires that all matrices in \eqref{seq_constirel} take the form $M_{ij} = M\delta_{ij} + M^H \epsilon_{ij}$. The non-dissipative part of the Josephson relation is fixed by the commutation relation \eqref{eq_com_2} and the Hamiltonian \eqref{eq_H0}, so that $\Omega^H = \omega_{\rm pk}$.

The background magnetic field breaks both parity ${\sf P}:(x,y)\to (-x,y)$ and time reversal ${\sf T}:t\to -t$, but preserves their product ${\sf PT}$. This symmetry imposes Onsager constraints on response functions. Specifically, one can show that for operators transforming under ${\sf PT}$ as $O_a\to \eta_{a}O_a$, with $\eta_{a} = \pm 1$, one has 
\begin{equation}\label{eq_ons}
G^R_{a b}(\omega) = 
	\eta_a \eta_b G^R_{b a} (\omega)\, .
\end{equation}
Once the hydrodynamic Green's functions have been computed --- we are about to review the method for doing this --- the Onsager constraint can be seen to fix $\widetilde \gamma = \gamma$ in \eqref{seq_constirel}. We impose this relation from this point on.

The retarded Green's functions are obtained by solving \eqref{seq_constirel} for the expectation values of the operators $\mathcal{O}_a = \{\varphi_i, j_i\}$ in terms of their sources $s_a = \{s_i,\, A_i\}$, where $A = E/i\omega$, as
\begin{equation}
G^R_{ab}(\omega) = \frac{\delta \langle \mathcal{O}_a(\omega)\rangle}{\delta s_b(\omega)}\, .
\end{equation}
This leads to the following Green's functions
\begin{subequations}\label{eq:allG}
\begin{align} \label{seq_GjjGff}
G^R_{j_ij_j}(\omega)
	&= i\omega \left[\hat\sigma^0 -\omega_{\rm pk} \frac{ \hat\gamma^2}{-i\omega + \hat\Omega}\right]_{ij}\, , \\
G^R_{\varphi_i\varphi_j}(\omega)
	& = -\frac{1}{\omega_{\rm pk}} \left[\frac{\hat\Omega}{-i\omega + \hat\Omega}\right]_{ij}\, , \\
G^R_{\varphi_i j_j}(\omega)
	= -G^R_{j_i \varphi_j}(\omega) & = i\omega\left[\frac{\hat\gamma}{-i\omega + \hat\Omega}\right]_{ij}\, , 
\end{align}
\end{subequations}
where the hats denote matrices. 

Positivity of dissipation implies that the anti-Hermitean part of the matrix $G^R_{ab}(\omega)$ must be positive definite (for example this imposes $\Im G^R_{aa}(\omega) \geq 0$ on diagonal entries). This leads to the following constraints
\begin{equation}
\sigma_0 \geq 0\, , \qquad\qquad
\Omega \geq 0\, , \qquad\qquad
\gamma^2 \leq \sigma_0 \frac{\Omega}{\omega_{\rm pk}}\, .
\end{equation}

The Kubo formulae for the dissipative and non-disspative parameters, given in \eqref{eq_Kubo} and \eqref{eq:gammaH} in the main text, including
\begin{subequations}
\begin{align}\label{eq_Kubo_Omega_app}
\frac{\Omega}{\omega_{\rm pk}}
	&= \lim_{\omega\to 0}\lim_{\epsilon\to 0}\frac{1}{\omega}\Im G^R_{\dot\varphi_x\dot\varphi_x}(\omega) \,, \\
\gamma\label{eq_Kubo_gamma_app}
	&
	= \lim_{\omega\to 0}\lim_{\epsilon\to 0}\frac{1}{\omega} \Im G^R_{\dot \varphi_x j_x}(\omega)\, ,\\
\gamma_H & = \lim_{\omega \to 0} \lim_{\epsilon\to 0}
        \Re G^R_{\varphi_y j_x}(\omega)\, ,
\end{align}
\end{subequations}
now follow from the explicit Green's functions in \eqref{eq:allG}, using
the identities
\begin{subequations}
\begin{align}
G^R_{\dot a b}(\omega)
	&= -i\omega G^R_{ab}(\omega) + \chi_{\dot a b}\, ,\\
G^R_{\dot a \dot b}(\omega)
	&= \omega^2 G^R_{ab}(\omega) + i\omega\chi_{\dot a b} + \chi_{\dot a \dot b}\, ,
\end{align}
\end{subequations}
and understanding the limit $\epsilon\to 0$ to mean that one only keeps the leading order term in relaxation $\Omega,\, \omega_p\sim \epsilon$. Note that the order of limits in the Kubo formula is essential in order to extract the correct result from \eqref{eq:allG}.

\section{Wigner solid hydrodynamics in a magnetic field}

An important objective in the main text was to develop a theory of the magnetophonon alone, decoupled from high energy modes.
Here we describe an extended hydrodynamics for translational order in a magnetic field, capturing both the magnetophonon and magnetoplasmon. This is only a sensible thing to do if the magnetoplasmon is below the local thermalization scale, which is not the case at large fields. Nonetheless, the hydrodynamic expressions provide a useful point of contact with more microscopic results. The formulae obtained may also be useful for weakly pinned translational order in small fields. The effect of a small magnetic field on a pre-existing pinned electron solid is to split the (coincident at $k=0)$ longitudinal and shear sound modes into a magnetoplasmon and a magnetophonon \cite{PhysRevB.18.6245}. 

The hydrodynamic variables are now the momenta $\pi_i$, the phases $\phi_i$ and the electric current $j_i$. The source conjugate to the momentum is the velocity $v_i$, the source conjugate to the phase is $s_i$ and the electric field is $E_i$. The key equations are the (approximate) conservation equation for the momentum, the phase-relaxed Josephson equation for the phase, and the constitutive relation for the current. The most general form these equations can take at leading (zeroth) order in gradients is
\begin{subequations}\label{seq_constirel_WC}
\begin{align}
\dot \pi \label{seq_consti_WC1}
	&= - \hat\Gamma \chi_{\pi\pi} v + \hat I s + \hat n E + \cdots\, , \\
\dot \phi
	&=  - \hat I v -\frac{\hat \Omega}{\omega_o^2} s - \hat \gamma E+ \cdots\, ,  \\ 
j  \label{seq_consti_WC3}
	&= \hat n v + \hat \gamma s + \hat \sigma_0 E+ \cdots\, .
\end{align}
\end{subequations}
Hatted variables are matrices. On the right hand we have only explicitly written the source terms (there are also fields on the right hand side, c.f. \eqref{seq_constirel}), as the dependence on the sources is sufficient to obtain the Green's functions.
The Onsager condition \eqref{eq_ons} was imposed and is responsible for the appearance of the matrices $\hat n$, $\hat\gamma$ and $\hat I$ in two equations each. The physical meaning of $\hat \Omega$ and $\hat \sigma_0$ is similar to in the main text, while $\hat \Gamma$ will be related to momentum relaxation. All matrices again have the form $\hat M= M\hat \delta + M^{\H} \hat\epsilon$ -- where $\hat \delta$ is the identity and $\hat \epsilon$ the Levi-Civita tensor -- with $M$ and $M^\H$ arbitrary, except for 
\begin{equation}
\hat I = \hat \delta + I^\H \hat \epsilon
\end{equation}
which follows from the normalization of the phason $[\phi_i(x), \pi_j] = i\delta_{ij}\delta^2(x)$. In the absence of a magnetic field, parity is preserved and all matrices are proportional to the identity -- in this case the constitutive relations reduce to those in \cite{PhysRevB.96.195128}.

The equations \eqref{seq_constirel_WC} describe the hydrodynamics of any system with spontaneously broken translations, without parity. One can easily obtain the conductivity from this theory:
\begin{equation}\label{eq_cond}
\hat\sigma
	= \hat \sigma_0 + \frac{\hat n^2}{\chi_{\pi\pi}}\frac{z+\hat \Omega - \omega_o^2\hat \gamma' (2 \hat I + \hat \gamma' (z + \hat\Gamma))}{(z+\hat\Gamma)(z+\hat\Omega) + \hat I^2 \omega_o^2}\, , 
\end{equation}
where $z=-i\omega$, and where we defined $\hat\gamma' =\chi_{\pi\pi} \hat \gamma /\hat n$ to simplify the expression.

When parity is broken specifically by a background magnetic field $B$, it is possible to express certain transport parameters appearing in \eqref{eq_cond} in terms of $B$. This is done by starting from \eqref{seq_constirel_WC} without parity breaking, and adding the Lorentz force term to the momentum (non)-conservation equation $\dot \pi_i = B \epsilon_{ij} j^j + \cdots$. Imposing consistency with the Onsager relations \eqref{eq_ons} then leads to
\begin{equation}\label{eq_relax_relation}
\hat n = n\hat \delta + B \hat\epsilon \hat \sigma_0\, , \qquad\qquad
\hat \Gamma = \Gamma\hat \delta + \frac{B^2}{\chi_{\pi\pi}} \hat \sigma_0 - \omega_c \hat \epsilon\, ,
\end{equation}
where $\omega_c = {nB}/{\chi_{\pi\pi}}$ is the cyclotron frequency. $\Gamma$ will be the momentum relaxation rate. Using (\ref{eq_relax_relation}) in (\ref{eq_cond}), the conductivity then has two finite frequency peaks, around $\omega_c$ and $\omega_o^2/\omega_c$, defining the magnetoplasmon and magnetophonon respectively.

The results in the main text can be recovered for this case (in which the magnetoplasmon is also hydrodynamical) by taking the limit $\omega_c\to \infty$, keeping $\omega_{\rm pk} = \omega_o^2/\omega_c$ and $\nu=n/B$ finite.
For illustrative purposes we will set $\hat\sigma_0,\, \hat\gamma = 0$; 
in general the $\omega_c \to \infty$ limit still maps on to the magnetophonon result without this simplification, but the map is more complicated. In the limit we then obtain
\begin{equation}
\hat\sigma(\omega) 
	=  \nu \hat \epsilon + \nu\omega_{\rm pk}\frac{\hat I^2}{z+\hat\Omega+\omega_{\rm pk}\hat \epsilon\hat I^2}
\end{equation}
which matches the magnetophonon conductivity \eqref{eq:www} or \eqref{seq_GjjGff} with replacements
\begin{equation}
\hat\sigma_0^{\rm mp}
	\rightarrow \nu \hat \epsilon \,, \qquad \qquad
a
	\rightarrow -I^H\, , \qquad \qquad
\hat \Omega^{\rm mp}
	\rightarrow \hat \Omega + \omega_{\rm pk} \hat \epsilon \hat I^2\, .
\end{equation}
%

\section{Screening of magnetophonons}

Long-range Coulomb interactions can significantly modify hydrodynamic correlation functions. See \cite{BEEKMAN20171} for an extended discussion of the effects of Coulomb interactions in translationally ordered states. As discussed in the text, screening should not be incorporated when computing the optical conductivity. However, other probes measure screened correlators or poles. These can be obtained by adding Coulomb interactions
\begin{equation}
H = \cdots + \int d^2 k \frac{n_k n_{-k}}{|k|}\, , 
\end{equation}
where here we assumed photons are not confined to the two dimensional plane (i.e. the photons are three dimensional). Resonances in response functions now will no longer be given by poles of $G^R_{nn}(\omega,k)$ but rather by solutions to
\begin{equation}\label{eq_screen}
|k| + G^R_{nn}(\omega,k) = 0 \,.
\end{equation}

The finite $k$ Green's function $G^R_{nn}(\omega,k)$ can be obtained for example by using the methods of appendix \ref{app_consti}. Only the non-dissipative Green's function is needed here. This can be obtained by analytic continuation of the Euclidean Green's function $G^E$ that follows from the effective action \eqref{eq_L}, extended to include coupling to the charge density (cf. also \eqref{eq_phidot} in the main text)
\begin{equation}\label{eq_L_phi_n}
\mathcal L_E
	= \epsilon^{ij} {\varphi_i}(i\d_\tau)\varphi_j -  \varphi_i \left(\kappa k^i k^j + \mu k^2\delta^{ij} \right)\varphi_j  + \frac{\gamma^H}{\chi} i k^i \varphi_i n + \frac{n^2}{2\chi}\, .
\end{equation}
Analytically continuing and removing a contact term $G^R(\omega) = G^E(i\omega) - \chi$ leads to the retarded Green's function
\begin{equation}
    G^R_{nn}(\omega,k)
        = \chi\frac{(\gamma_H^2/\chi)\mu k^4}{\omega^2 - \mu(\kappa+\mu+(\gamma_H^2/\chi))k^4}\, .
\end{equation}
Solving \eqref{eq_screen} and using $\gamma_H = \sqrt{\nu}$ then gives the well-known dispersion relation of a screened magnetophonon \cite{PhysRevB.46.3920}
\begin{equation}\label{eq_screenedpole}
\omega
	= \pm \sqrt{\mu \nu} \, k^{3/2}\, .
\end{equation}

\end{document}